\RequirePackage[2020-02-02]{latexrelease}
\documentclass[%
 aip,
 amsmath,amssymb,
 reprint,%
]{revtex4-1}
\usepackage{graphicx}
\usepackage{dcolumn}
\usepackage{bm}

\usepackage{mathptmx}
\usepackage{etoolbox}

\usepackage{float}
\usepackage{graphicx}
\usepackage{dcolumn}
\usepackage{bm}

\usepackage[bottom]{footmisc}
\usepackage{xcolor}
\usepackage{stmaryrd}

\usepackage{amsmath, bm}
\usepackage{amssymb}
\usepackage{dsfont}
\usepackage{stmaryrd}
\usepackage{mathptmx}
\usepackage{xcolor}
\usepackage{mdframed}
\usepackage{graphicx}
\usepackage{epsfig}
\DeclareMathAlphabet{\mathbbmsl}{U}{bbm}{m}{sl}

\newcommand{\beq}{\begin{equation}}
\newcommand{\bseqs}{\begin{subequations}}
\newcommand{\eseqs}{\end{subequations}}
\newcommand{\balign}{\begin{align}}
\newcommand{\ealign}{\end{align}}
\newcommand{\eeq}{\end{equation}}
\newcommand{\beql}{\begin{equation} \label}
\newcommand{\beqs}{\begin{eqnarray}}
\newcommand{\eeqs}{\end{eqnarray}}
\newcommand{\beas}{\begin{eqnarray*}}
\newcommand{\eeas}{\end{eqnarray*}}
\newcommand{\ber}{\begin{array}}
\newcommand{\eer}{\end{array}}
\newcommand{\becs}{\begin{cases}}
\newcommand{\eecs}{\end{cases}}

\newcommand{\rz}{{\mathbb{R}}}

\newcommand{\bfa}{{\mathbf a}}

\newcommand{\bfc}{{\mathbf c}}

\newcommand{\bfe}{{\mathbf e}}
\newcommand{\bff}{{\mathbf f}}
\newcommand{\bfg}{{\mathbf g}}

\newcommand{\bfi}{{\mathbf i}}
\newcommand{\bfj}{{\mathbf j}}
\newcommand{\bfk}{{\mathbf k}}

\newcommand{\bfp}{{\mathbf p}}
\newcommand{\bfq}{{\mathbf q}}

\newcommand{\bfv}{{\mathbf v}}

\newcommand{\bfx}{{\mathbf x}}

\newcommand{\bfD}{{\mathbf D}}

\newcommand{\bfI}{{\mathbf I}}

\newcommand{\bfL}{{\mathbf L}}
\newcommand{\bfM}{{\mathbf M}}

\newcommand{\bfP}{{\mathbf P}}
\newcommand{\bfQ}{{\mathbf Q}}
\newcommand{\bfR}{{\mathbf R}}

\newcommand{\bfT}{{\mathbf T}}

\newcommand{\bfW}{{\mathbf W}}

\newcommand{\bbm}{\begin{bmatrix}}
\newcommand{\ebm}{\end{bmatrix}}
\newcommand{\bfalpha}{{{\boldsymbol{\alpha}}}}
\newcommand{\bfbeta}{{{\boldsymbol{\beta}}}}

\newcommand{\sgn}{{\rm sgn}}

\newcommand{\omegahat}{{\hat{\omega}}}

\newcommand{\bfsigma}{{{\boldsymbol{\sigma}}}}

\newcommand{\bfomega}{{{\boldsymbol{\omega}}}}

\newcommand{\bfsigmahat}{{\hat{\boldsymbol{\sigma}}}}

\newcommand{\Id}{{\rm Id}}

\newcommand{\bftau}{{{\bf \tau}}}

\newcommand{\half}{\frac{\;1}{\;2}}

\newcommand{\dt}{d {\bf t}}

\newcommand{\aand}{{{\rm\;\; and\;\;}}}

\newcommand{\Tr}{{{\rm Tr}}}

\newcommand{\aas}{{{\rm\; as\;}}}

\newcommand{\bfxtld}{{\tilde{\bfx}}}

\newcommand{\sym}{{\rm sym}}

\newcommand{\grad}{{\rm grad}}

\newcommand{\eff}{{\rm eff}}

\renewcommand{\tt}{{\rm tt}}
\newcommand{\tr}{{\rm tr}}
\newcommand{\rr}{{\rm rr}}

\newcommand{\SO}{{\rm SO}}
\newcommand{\bfRhat}{{\hat{\bf R}}}
\newcommand{\bfIhat}{{\hat{\bf I}}}
\renewcommand{\dt}{{\vartriangle\! \! t}}
\newcommand{\bfomegahat}{{\hat{\boldsymbol \omega}}}
\newcommand{\bfTheta}{{\bm{\Theta}}}
\newcommand{\bfvdot}{{\dot{\bfv}}}

\newcommand{\bfqbar}{\overline{\bfq}}

\newcommand{\Mhat}{\hat{M}}
\newcommand{\Upsilonhat}{\hat{\Upsilon}}

\makeatletter
\def\@email#1#2{%
 \endgroup
 \patchcmd{\titleblock@produce}
  {\frontmatter@RRAPformat}
  {\frontmatter@RRAPformat{\produce@RRAP{*#1\href{mailto:#2}{#2}}}\frontmatter@RRAPformat}
  {}{}
}%
\makeatother

\begin{document}

\preprint{AIP/123-QED}



\title[]{Diffusion of microstructured anisotropic particles in an external field}
\author{Tianyu Yuan}
\affiliation{Institute for Advanced Study, Chengdu University, Chengdu, Sichuan 610106, P.R. China}
\affiliation{State Key Laboratory for Turbulence and Complex Systems, Department of Mechanics and Engineering Science, College of Engineering, Peking University, Beijing 100871, P.R. China}

\author{Liping Liu}
\affiliation{Department of Mechanical and Aerospace Engineering, Rutgers University, Piscataway, New Jersey 08854, USA}
\affiliation{Department of Mathematics, Rutgers University, Piscataway, New Jersey 08854, USA}
\email{liu.liping@rutgers.edu. Corresponding author}

\author{Jianxiang Wang}
\affiliation{State Key Laboratory for Turbulence and Complex Systems, Department of Mechanics and Engineering Science, College of Engineering, Peking University, Beijing 100871, P.R. China}
\affiliation{CAPT-HEDPS, and IFSA Collaborative Innovation center of MoE, College of Engineering,Peking University, Beijing 100871, P.R. China}

\date{\today}

\begin{abstract}
Microstructured particles are widely used in   industries and state-of-the-art  research and development. Diffusion of  particles, particularly, controlled diffusion   by a remotely applied   field,   has  inspired novel applications ranging from targeted drug deliveries, novel procedures  for quantifying physical properties of nanoparticles and ambient fluids, to  fabrication of composites with enhanced properties. In this work, we report a systematic analysis on   field-controlled diffusion of microstructured particles. In account of  shape anisotropy and structural heterogeneity  of a  particle, we study   coupled Brownian motions of the particle in $\rz^3\times \SO(3)$.  Starting from the microscopic stochastic differential equations  of motions, we  achieve the coarse-grained  Fokker-Planck equation that governs  the evolution of the probability distribution function with respect to the position and  orientation of the particle. Under some mild conditions, we identify the   long-time diffusivity for    microstructured particles in an  external field.  The formulation  is applicable to  microstructured particles of arbitrary shapes and   heterogeneities. As examples of applications, we analyze the diffusion  of a heterogeneous spheroidal particle and  a pair of spheroidal particles bonded by an elastic ligament. For heterogeneous spheroidal particles, we obtain explicit generalized Stokes-Einstein's relations for diffusivity that accounts for the effects of shape anisotropy, heterogeneity, and an external alignment field. For pairs of spheroidal particles, we consider the superimposed relaxation process    from an initial non-equilibrium state to the final equilibrium state. The   anomalous scaling of Mean Square Displacement (MSD) with respect to time of such processes    may   provide important insight for understanding anomalous diffusions observed in migration of macromolecules and  cells in  complex viscoelastic media. \end{abstract}

\maketitle

\section{Introduction}

Brownian motion of  particles was first observed and described  in 1827 by Brown. This random, uncontrolled motion of  particles in  a fluid, driven by the bombardment of ambient fluid molecules, gives rise to the macroscopic phenomenon of diffusion \citep{Einstein1905On}    and convincing evidence of atomistic theory of matter. \citep{Perrin1908Les}  These seminal works of Einstein and Perrin entail a fundamental understanding of Brownian motions of rigid spherical particles in a Newtonian fluid. In particular,   the intimate relation is now well understood between the seemingly uncontrolled microscopic random motions of microparticles and the macroscopic diffusion   of a solute in a solvent, i.e.,  the upscaling from the microscopic processes to   macroscopic properties.

Nowadays, many advanced  technologies necessitate non-standard Brownian  processes of microstructured particles in a complex medium.   For instance,   drug   release from  swellable polymer    involves   Brownian motions in a porous elastic medium, giving rise to non-Fickian diffusion behavior.~\citep{ende1995}  Also, the diffusion of  drug molecules in the tumor interstitium~\citep{Jain1987tumor} and the diffusion of univalent cations and anions out of liquid crystals of lecithin~\citep{Bangham1965ion}  require consideration of viscoelasticity and anisotropy of the medium and multiphysical interactions between the particle and   ambient medium. These non-standard  Brownian  processes demand a more comprehensive analysis of motions of microstructured particles  in a complex   medium and its ramification in macroscopic anomalous diffusions.

 With the   advancement of experimental techniques including  fluorescence recovery and confocal microscopy, a great number of experiments   have been made to  Brownian motions of various microstructured  particles in a complex medium.   Perrin~\citep{Perrin1934Mouvement,Perrin1936Mouvement} has observed the Brownian motion of ellipsoidal particle as early as in 1930's.
 Han {\em et al.}\citep{Han2006Brownian} analyzed and measured the two dimensional Brownian motions of ellipsoidal particles in water, which was later extended to  quasi-two-dimension  to account for the effects of shape anisotropy and confinement. \citep{Han2009Quasi} In the field of nanotechnology, the Brownian motions of a variety of nano-particles are visualized and measured, including
 single-walled carbon nanotubes, \citep{Duggal2006carbonnanotube} copper oxide nanorods, \citep{Cheong2010Rotational} carbon nanofibers, \citep{Bhaduri2008carbonnanofiber} etc. The diffusion behavior of   microstructured particles or microorganisms    have   been characterized, ranging from colloidal trimer, \citep{Kraft2013colloidal} graphene flakes, \citep{Marago2010Brownian} boomerang particles, \citep{Chakrabarty2014Brownian} actin filaments, \citep{Koster2005Brownian} to Leptospira interrogans.~\citep{Koens2014Leptospira}

Conventional analysis of Brownian motions starts from a spherical rigid  homogeneous particle in a   Newtonian fluid.  
The central result of Einstein\citep{Einstein1905On} and Perrin~\citep{Perrin1908Les}  is termed as the Stokes-Einstein's relation: $D=\mu k_B T$, where  $D$, $\mu$, $k_B$, and $T$ are the macroscopic diffusivity, microscopic mobility,  Boltzmann's constant, and absolute temperature, respectively. Combined with Stokes's formula for the mobility, the Stokes-Einstein's  relation provides early measurement of the Boltzmann's constant  (or Avogadro constant) and decisive evidence of atomistic theory of matter.  A microstructured particle, however, could  have heterogeneous density  and complex shape. From a microscopic viewpoint, the hydrodynamic interaction between   the microstructured particle and ambient fluid can no longer be sufficiently captured by a single scalar   mobility $\mu$. Nevertheless, at the absence of external field  we expect   particles would diffuse isotropically with   $D=\langle x(t)^2\rangle/2t$ if the time scale $t$ is long enough to smear out all microscopic short-time rotations of the particles. From this viewpoint, we expect two distinct regions of diffusion behaviors delimited by  the {\em crossover time} scale $t_{\rm cross}$.~\citep{Han2006Brownian}   On a short time-scale $t<t_{\rm cross}$,   the initial orientation of the anisotropic particle has a significant influence on movements of the particle, resulting in an anisotropic and time-dependent diffusivity tensor $\bfD\sim \langle \bfx(t)\otimes \bfx(t)\rangle/2t$. On the other hand,  if the time-scale $t \gg t_{\rm cross}$,  the classic Stokes-Einstein's relation should   hold  for some ``effective'' or ``average'' mobility $\mu^\eff$, provided that the ambient fluid is isotropic and there is no external field breaking the rotational symmetry.

In an effort to quantify the crossover time scale and generalize the Stokes-Einstein's relation, Brenner~\cite{Brenner1964extension, Brenner1967Coupling}  conducted a systematic study on the   Brownian motions of particles of arbitrary shapes. His analysis revealed  two important effects of shape anisotropy: (i) the Stokes' formula for mobility of a spherical particle shall be replaced by an anisotropic mobility tensor for a  particle of general shape, and (ii) the {\em hydrodynamic center}, i.e., the point of action of the resultant hydrodynamic forces on the non-rotating particle, in general may not coincide with the center of mass of the particle. For particles with hydrodynamic center coinciding with  the center of mass,~\citep{Brenner1964extension, Bernal1980, Harvey1980} the translational and rotational motions are uncoupled and solutions to the associated stochastic differential equations are manageable.~\citep{Han2006Brownian, Cheong2010Rotational, Chakrabarty2014Brownian, Yuan2021} For particles of general shapes and heterogeneities,  the translational and rotational degrees of freedom  are intrinsically coupled, besides the technical difficulties arising from the stochastic forces of thermal agitations and nonlocality of hydrodynamic forces on the particle. A major theme of this work is to analyze this  coupled stochastic  equations of motions, quantify the crossover time scale $t_{\rm cross}$, and achieve generalized Stokes-Einstein's relations for  the long-time diffusivity   in terms of geometrical and physical properties of the particle and ambient fluid.

A second theme of our current work is   to explore the effects of external  fields on the diffusivity of  microstructured particles.    External electric or magnetic field  can be remotely applied to  suspensions and easily controlled.  From a non-stochastic viewpoint, external fields have two major effects on microstructured particles: (i) suspension particles in general move along field lines to either concentrate to (or dilate from)   regions of larger field strength if the field is nonuniform, and (ii) suspension particles tend to align with field lines.  The movements driven by external fields, though inevitably randomized by thermal agitations of ambient fluid molecules, are expected to have an overall effect on the cross-over time scale and the long-time diffusivity. The ability of manipulating  the motion or diffusion of   particles in suspension  has  inspired many   applications. For instance, an electric field can assist the scalable  fabrication of  composites incorporating vertically-aligned carbon nanotubes  \citep{Castellano2015Electrokintics} or well-organized boron-nitride nanotubes.~ \citep{Cetindag2017Surface}   A magnetic field can be used to tune the acoustic attenuation of the suspensions  of subwavelength-sized nickel particles.~  \citep{Yuan2017Tunable} Further, the field-controlled  motion or diffusion of  particles in suspension paves  new  anvenues for applications in   induced-charge electrophoresis,~\citep{Squires2006Breaking}   contactless characterization of the semiconductor,~ \citep{Yuan2019Contactless}   separation of the particle,~ \citep{Doiparticlesorting} etc.

To understand the field-controlled diffusion of particles, Grima et al. \citep{Grima2007Brownian} studied the quasi-two-dimensional Brownian motion of an ellipsoidal rigid particle in the electrophoresis and microconfinement and found that the long-time diffusivity of an asymmetric particle  is anisotropic in the presence of external forces and  depends  on the shape of the particle. Gu\"{e}ll {\em et al.}\citep{Guell2010Anisotropic}  studied the diffusion of an ellipsoidal particle in an external rotating magnetic field and showed the correlation of the crossover time scale with the frequency and amplitude of the field. Aurell {\em et al.}\citep{Aurell2016Diffusion} used a systematic multi-scale technique to study the diffusion of an ellipsoid under the application of   a constant external force and found that the diffusivity parallel to the direction of the force field is $\frac{4}{3}$ of that perpendicular to the direction the force field. Obasanjo \citep{Obasanjo2016Response} experimentally studied the electro-rotation and electro-orientation of the particle  and  measured the electric-field-tuned diffusion coefficients of ellipsoidal particles in the AC field.   Segovia-Guti\'{e}rrez {\em et al.}  \citep{Segovia2019Diffusion} experimentally studied the rotational and translational dynamics of trimers in a quasi-two-dimensional system under a random light field and found the coupling between the rotational motion and the translational motion of trimers relies on the length of the scale of the particle. In our prior work,~\citep{Yuan2021} we studied the diffusion of carbon nanotube under an aligning AC electric field,  gave an explicit formula to the anisotropic diffusivity tensor, and experimentally validated the formula. In particular, we  found that the diffusion coefficient parallel to the field increases and the diffusion coefficient  perpendicular to the field decreases as the field strength increases. The trace of diffusivity tensor tensor remains as constant. In all these works,  the hydrodynamic center of the particle coincides with the center of mass  because of axis-symmetry, and hence the translational and rotational motions are essentially decoupled.

 In the this  work we  focus on   the long-time diffusivity  of   microstructured particles of general shapes and heterogeneities in an external field. Starting from the  microscopic   stochastic equations of motions with coupled translational and rotational degrees of freedom,   we find the coarse-grained Fokker-Planck equation that governs the evolution of the Probability Distribution Function (PDF)    in the configurational space consisting of position, orientation, linear velocity and angular velocity of the particle. From the classical Boltzmann's distribution for equilibrium states, we determine the fluctuation coefficients associated with Brownian forces. Next, we neglect the effects of inertia and consider a Lagevin-type equation for the position and orientation of the particle. Again by studying the associated Fokker-Planck equation, we extract the macroscopic long-time diffusivity in terms of properties of microstructured particles and ambient fluid. This approach is versatile and self-contained and could be applied to study more general diffusion phenomena, e.g., deformable particles in a visco-elastic ambient medium.

On the technical side, the generalized Stokes-Einstein's relations we obtain for the long-time diffusivity of general microstructured particles  involve evaluations of the mobility tensors    and integrals over the continuous group of all spatial orientations (or rigid rotations) $\SO(3)$. The former concerns the classical Stokes' flow around a particle of arbitrary shape with certain linear velocity and angular velocity and has been addressed in a number of earlier works.~\cite{Brenner1964extension, Brenner1967Coupling} In particular, explicit solutions to the mobility tensor  are available for ellipsoids.~\cite{Brenner1964extension, Brenner1967Coupling}  Integrations over the continuous group $\SO(3)$ are achieved by  a special parametrization of $\SO(3)$ that is closely related to  the quaternion algebras.\citep{Quaternions}   Upon employing approximations based on the mobility tensor of ellipsoids and quaternion algebras, we achieve explicit formulas for long-time diffusivitives of  general heterogeneous ellipsoids and two spheroids bonded by some ligaments.  These results could be further applied to   improve the design and functionality of microstructured particles in colloidal systems and  create fundamental understanding of the anomalous diffusion of macromolecules in complex media.

The paper is organized as follows.  In Sec.~\ref{sec:EOM} we present  our approach from microscopic equations of motions for Brownian motions to Fokker-Planck equations for macroscopic  evolution of PDF in configurational space.  In Sec.~\ref{sec:Laginvin} we analyze the Fokker-Planck equations, quantify the   crossover time scale, and obtain a formula for the long-time diffusivity of   microstructured particles under an external field.   In Sec.~\ref{Sec:app} we present explicit  results for the long-time diffusivity of   heterogeneous spheroids, which generalizes the classical Stokes-Einstein's formula to account for  the effects of shape anisotropy and heterogeneities.   Based on these explicit results, we then study the diffusion of a pair of spheroids   undergoing a transition  from an initial non-equilibrium state to the final equilibrium state. The non-standard scaling behavior of MSD versus time of these processes may be a good model for  understanding the anomalous diffusions, e.g., migration of macromolecules and  cells in the complex medium.

\noindent {\em Notation.}\;  Direct notation is employed for brevity and  transparency  of physical interpretation whenever possible. {  Frequently, recognizing that many readers may be more familiar with the index notation, we also presented translations in index form to illustrate details of the calculations.} Vectors and tensors are denoted by bold symbols such as $\bfx, \bfv, \bfQ$, etc., while scalars are denoted by $\psi, \eta$, etc. For a vector field $\bfv$, in index form the gradient operator $\grad \bfv$ is equivalent to $(\bfv)_{i, j}$ with $i$ $(j)$ being the first (second) index.
 When index notation is  in use, the convention of summation over repeated index are followed.

 \section{Equation of motion} \label{sec:EOM}

We consider the  Brownian motion of a  particle in a Newtonian fluid  under the application of some   external field. Denote by $\bfx$  the position of the center of mass with respect to a fixed global frame $\{\bfe_i: i=1,2,3\}$,   $\bfv=\dot{\bfx}(t) $ is  the velocity of the center of mass of the particle,  (``$\cdot$''  denotes $\frac{\mathrm{d}}{\mathrm{d} t}$) $\bfomega$    the angular velocity of the particle, and
$m$ (resp. $\bfI\in \rz^{3\times 3}_\sym$)   the mass (resp. the  moment of inertia with respect to the center of mass and the global frame).  The presence of an external field in general exerts certain external force ($\bfg^{e} \in\mathbb{R}^3$) and torque
($ {\bftau}^e\in\mathbb{R}^3$) on the microparticle.  In addition, the bombardments of ambient molecules on the particle give rise to a random force  $\bfg^{B}$ and random torque ${\bftau}^B $.

Suppose the particle moves in the fluid with linear velocity $\bfv$ and angular velocity $\bfomega$. The particle   generates a  flow in the ambient fluid and consequently, suffers from a drag force $\bfg^d$ and torque $\bftau^d$  because of the viscosity of the fluid. Since the Reynold number is small, it suffices to consider Stokes' flow and conclude that the force $\bfg^d$ and torque $\bftau^d$  on the particle are related with the   linear velocity $\bfv$ and angular velocity $\bfomega$ by a linear transformation:
\beqs\label{frictiontotal}
\begin{bmatrix}
\bfg^d\\
\bftau^d
\end{bmatrix}=-\bfR
\begin{bmatrix}
\bfv\\
\bfomega
\end{bmatrix},
\eeqs
where the   drag   matrix  $\bfR\in \rz^{6\times 6}_\sym$ is symmetric and positive-definite. For future convenience, we write the drag matrix in a block form as
\beqs \label{eq:bfRblock}
\begin{split}
\bfR=
\begin{bmatrix}
    \bfR^\tt  &\bfR^\tr  \\
   ( \bfR^\tr)^T &\bfR^\rr \\
    \end{bmatrix} ,
    \end{split}
\eeqs
where    $\bfR^{(\tt, \rr, \tr)}\in\rz ^{3\times3}$ pertains to the translational motion, the rotational motions, and the coupling  between translational and rotational motions, respectively.

  From the classic rigid-body mechanics, the equation of motion for the particle  can be expressed as
\beqs\label{eq:kinetic anisotropic000}
\begin{split}
&\frac{\mathrm{d}}{\mathrm{d} t}( m\bfv) =-{\bf R}^\tt  \bfv -{\bf R}^\tr \bfomega+\bfg^e +\bfg^B,\\
 &\frac{\mathrm{d}}{\mathrm{d} t} (\bfI \dot{\bfomega})= - {\bfR}^\rr \bfomega-{\bfR}^{\tr^T}\bfv+{\bftau}^e+{\bftau}^B.
\end{split}
\eeqs
For  an orthonormal body frame $\{\bff_i: i=1,2,3\}$ fixed on the particle,  let $\bfQ(t)=\bfe_i\otimes\bff_i(t)\in\SO(3)$ be the associated orthogonal  matrix
and
\beas
\begin{split}
& (\bfRhat^{(\tt, \tr, \rr)}, \bfIhat, \bfsigmahat_r)=\bfQ(\bfR^{(\tt, \tr, \rr)}, \bfI, \bfsigma_r)\bfQ^T ,\\
& (\hat{\bfomega},  \hat{\bftau}^e,\hat{\bftau}^B)=\bfQ({\bfomega},  {\bftau}^e,{\bftau}^B)
 \end{split}
\eeas
  the tensors and vectors  with respect to the body frame $\{\bff_i: i=1,2,3\}$.  Then the equation of motion~\eqref{eq:kinetic anisotropic000}  can be rewritten as~\citep{Yuan2021}
\beqs\label{eq:kinetic anisotropic}
\begin{split}
&m\dot{\bfv}=-\bfQ^T\hat{\bf R}^\tt \bfQ   \bfv -\bfQ^T\hat{\bf R}^\tr \bfomegahat+\bfg^e +\bfg^B,\\
 &\bfIhat \dot{\bfomegahat}= - \hat{\bfR}^\rr \omegahat- \hat{\bfR}^{\tr^T} \bfQ \bfv+\hat{\bftau}^e+\hat{\bftau}^B,
\end{split}
\eeqs
where the term $ \bfomegahat\times \bfIhat\bfomegahat$ has been neglected  as compared with $\bfIhat \dot{\bfomegahat}$.~\citep{Yuan2021}
  We remark that  the drag matrices $\bfRhat^{(\tt, \tr, \rr)}$ and moment of inertia $\bfIhat$ with respect to the body frame are independent of the time or the motion of the particle. They  are geometrical and physical properties of the particle  and the ambient fluid. In this work we will   focus on  the case that the translational and  rotational  motions of the particle  are intrinsically coupled in the sense that $\bfRhat^\tr\neq{\bf 0}$.

We now determine the random force $\bfg^B$ and torque $\hat{\bftau}^B$    from thermal agitations. Presumably, these random force and torque on the particle is independent of the external fields. Therefore, for the purpose of fixing $\bfg^B$ and $\hat{\bftau}^B$ it is convenient to  analyze   \eqref{eq:kinetic anisotropic}   at the absence of the external field, i.e.,
\beqs\label{eq: for velocity}
\begin{split}
\dot{\bfv}&=-\frac{1}{m}\bfQ^T\hat{\bf R}^\tt \bfQ   \bfv -\frac{1}{m}\bfQ^T\hat{\bf R}^\tr \bfomegahat +\frac{1}{m}\bfg^B,\\
 \dot{\bfomegahat}&= -  \bfIhat^{-1}\hat{\bfR}^\rr \omegahat- \bfIhat^{-1}\hat{\bfR}^{\tr^T}\bfQ \bfv+\bfIhat^{-1}\hat{\bftau}^B.\\
\end{split}
\eeqs
To proceed, we assume that
the generalized random force   can be expressed as
\beqs\label{eq: sigmatildeexpression}
 \begin{bmatrix}\frac{1}{m}\bfg^B\\\bfIhat^{-1}\hat{\bftau}^B\end{bmatrix}=\tilde{\bm{\sigma}}\tilde{\bm{\xi}},
\eeqs
where $\tilde{\bm{\sigma}}\in \rz^{6\times 6}$ is called the fluctuation coefficients, and $\tilde{\bm{\xi}}(t) $ represents normalized six-dimensional uncorrelated white noises. In other words,  the process $\tilde{\bm{\xi}}(t) $ satisfies
\beqs \label{eq:xidef}
\langle \tilde{\xi}(t)\rangle =0,\qquad
\langle \tilde{\xi}_i(t) \tilde{\xi}_j(t')\rangle=\delta_{ij}\delta(t-t'),
\eeqs
where $\delta_{ij}$ is the Kronecker delta, and $\delta$ is the Dirac function.
  Next, we introduce notation
\beqs\label{eq: l}
\tilde{\bfv}:=\begin{bmatrix}\bfv\\\bfomegahat\end{bmatrix},\qquad
\bfL:=\begin{bmatrix}-\frac{1}{m}\bfQ^T\hat{\bf R}^\tt \bfQ & -\frac{1}{m}\bfQ^T\hat{\bf R}^\tr \\- \bfIhat^{-1}\hat{\bfR}^{\tr^T}\bfQ&- \bfIhat^{-1}\hat{\bfR}^\rr \end{bmatrix}.
\eeqs
 By \eqref{eq: l} and \eqref{eq: sigmatildeexpression},  we  rewrite \eqref{eq: for velocity} in a compact form as \beqs\label{governingequation}
\dot{\tilde{\bfv}}=\bfL\tilde{\bfv}+\tilde{\bm{\sigma}}\tilde{\bm{\xi}},
\eeqs
which may be recognized as  a   stochastic   differential equation (SDE) for  the six-dimensional generalized velocity $\tilde{\bfv}$.

To fix the fluctuation coefficients $\tilde{\bm{\sigma}}$, we consider the probability distribution function  (PDF)  $P=P(\tilde{\bfv}, t)$ in the generalized velocity  space.  Physically, the quantity $P(\tilde{\bfv}, t)\mathrm{d}\tilde{\bfv}$ is the probability of  finding the particle  with generalized velocity from the infinitesimal  element $\mathrm{d}\tilde{\bfv}$ centered at $\tilde{\bfv}$. 
From the master equation, it can be shown that the   PDF $P(\tilde{\bfv}, t)$ associated with the stochastic process~\eqref{governingequation}  satisfies the Fokker-Planck equation:~\citep{Kampen2007Stochastic}
\begin{equation} \label{Fokker-Planck for rtilde}
\frac{\partial P(\tilde{\bfv}, t)}{\partial t}=\nabla_{\tilde{\bfv}} \cdot \Big\{-\tilde{\bfalpha} P(\tilde{\bfv}, t)+\frac{1}{2}\tilde{\bfbeta}\nabla_{\tilde{\bfv}}  P(\tilde{\bfv}, t)\Big\},
\end{equation}
where
\beqs\label{eq: alphabeta}
\label{alphabeta}
\begin{split}
&\boldsymbol{\tilde{\alpha}}=\lim_{\dt \rightarrow 0} \frac{\langle \tilde{\bfv}(t+\dt )-\tilde{\bfv}(t)\rangle }{\dt },\\
&\boldsymbol{\tilde{\beta}}=\lim_{\dt \rightarrow 0} \frac{\langle  [\tilde{\bfv}(t+\dt )- \tilde{\bfv}(t)] \otimes [\tilde{\bfv}(t+\dt )- \tilde{\bfv}(t)]\rangle }{\dt }.
\end{split}
\eeqs
By  \eqref{governingequation} and \eqref{eq: alphabeta},   it can be shown  that~\cite{evans2012stochastic, Yuan2021}
\beqs\label{eq: alphabetatilde}
\begin{split}
&\boldsymbol{\tilde{\alpha}}=\bfL\tilde{\bfv}\quad\mathrm{and}\quad\boldsymbol{\tilde{\beta}}=\tilde{\bm{\sigma}}\tilde{\bm{\sigma}}^T.
\end{split}
\eeqs

From the classical statistical physics, the Maxwell-Boltzmann distribution
\beqs\label{stationary}
 P^s(\tilde{\bfv})\propto \exp{\left[-\left(\frac{m|\bfv|^2}{2k_BT}+\frac{\bfomegahat\cdot \bfIhat\bfomegahat}{2k_BT}\right)\right]}
 \eeqs
should be the stationary solution to \eqref{Fokker-Planck for rtilde}. Inserting \eqref{stationary} and \eqref{eq: alphabetatilde} into \eqref{Fokker-Planck for rtilde}, we find that
\beqs\label{fluctuationsdetermined}
\tilde{\bm{\sigma}}\tilde{\bm{\sigma}}^T=2k_BT\begin{bmatrix}\frac{1}{m^2}\bfQ^T\hat{\bf R}^\tt \bfQ&\frac{1}{m}\bfQ^T\hat{\bf R}^\tr\bfIhat^{-1}\\\frac{1}{m}\bfIhat^{-1}\hat{\bf R}^{\tr^T}\bfQ&\bfIhat^{-1}\hat{\bf R}^\rr\bfIhat^{-1}\end{bmatrix},
\eeqs
which essentially specify   the random force $\bfg^B$ and torque $\hat{\bftau}^B$ in  \eqref{eq:kinetic anisotropic}.\\

\section{The long-time diffusivity of   microstructured particles}  \label{sec:Laginvin}

We now consider the diffusion of particles in space.  Since the Reynolds number is   low, it is widely accepted that  the inertia terms (i.e. $m\bfvdot$ and $\bfIhat \dot{\bfomegahat}$) in \eqref{eq:kinetic anisotropic} could be neglected for   the evolution  of particles in configurational space of spatial position and orientation, which is often referred to as  the over-damped theory.~\citep{Travitzoverdamped} If $\bfRhat^\tr\equiv 0$, it is clear that the first of~\eqref{eq:kinetic anisotropic}  is   independent of the second of  \eqref{eq:kinetic anisotropic},  and hence the translational diffusivity can be determined without considering rotational motions.

To account for the case of  $\bfRhat^\tr\neq 0$, we first notice that the angular velocity $\bfomegahat$ and rotation matrix $\bfQ$ are kinematically related in   \eqref{eq:kinetic anisotropic}.  Specifically, the angular velocity $\bfomegahat$, as the (pseudo-)vector associated with the skew-symmetric matrix $\bfQ\dot{\bfQ}^T$, must satisfy
\begin{equation}  \label{Qdot}
\bfQ\dot{\bfQ}^T\bfa=\bfomegahat\times \bfa,\quad\forall \bfa\in\rz^3,
\end{equation}
where the rotation matrix $\bfQ$  represents the orientation of the particle with respect to the global frame $\{\bfe_i: i=1,2, 3\}$. For future calculations, it is necessary to introduce some parametrization $\boldsymbol{\Theta}\in\mathbb{R}^{m}$   for the rigid rotations in  the continuous group $\SO(3)$, e.g.,  the  quaternion ($m=4$) or the Euler angles (m=$3$). Once the parametrization $\boldsymbol{\Theta}$ is chosen,   a transformation matrix $\bfT\in\mathbb{R}^{m\times 3}$ can be introduced to relate the angular velocity $\bfomegahat$ and the rate of the change of $\bfTheta$, i.e.,
\beqs\label{omegaTrelation}
 \dot{\boldsymbol{\Theta}}=\bfT\bfomegahat.
 \eeqs
Next, we introduce the position-orientation coordinates
\beqs\label{tildex}
\tilde{\bfx}=\begin{bmatrix}\bfx\\\boldsymbol{\Theta}\end{bmatrix}\in\mathbb{R}^{3+m}.
\eeqs
By \eqref{omegaTrelation}  the position-orientation coordinates $\tilde{\bfx}$ and generalized velocities $\tilde{\bfv}$ in~\eqref{eq: l}  are related by
\beqs\label{eq: Ptensor}
\dot{\tilde{\bfx}}=\bfP\tilde{\bfv}
 \eeqs
where
 \beqs\label{eq: Ptensor2}
 \bfP=\begin{bmatrix}\Id&0\\0&\bfT\end{bmatrix}
  \eeqs
with  $\Id\in\mathbb{R}^{3\times3}$ denoting the identity tensor.

Based on equations \eqref{tildex}-\eqref{eq: Ptensor2} and the over-damped postulate, we rewrite \eqref{eq:kinetic anisotropic} as
\beqs\label{governingequationnew2}
 \dot{\tilde{\bfx}}=-\bm{\tilde{L}}\tilde{\bfg}-\bm{\tilde{L}}\tilde{\bm{\sigma}}^*\tilde{\bm{\xi}},
 \eeqs
where
 \begin{widetext}
\beqs\label{haha}
\begin{split}
\bm{\tilde{L}}:=&\begin{bmatrix}
 -\left(\bfR^\tt-\bfR^\tr\bfR^{\rr^{-1}}\bfR^{\tr^T}\right)^{-1}
 &\left(\bfR^\tt-\bfR^\tr\bfR^{\rr^{-1}}\bfR^{\tr^T}\right)^{-1}\bfR^{\tr}\bfR^{\rr^{-1}}\bfQ^{T}\\
 \bfT\bfQ\left(\bfR^\rr-\bfR^{\tr^T}\bfR^{\tt^{-1}}\bfR^\tr\right)^{-1}\bfR^{\tr^T}\bfR^{\tt^{-1}}
 &-\bfT\bfQ\left(\bfR^\rr-\bfR^{\tr^T}\bfR^{\tt^{-1}}\bfR^\tr\right)^{-1}\bfQ^T
 \end{bmatrix}, \qquad  \qquad \tilde{\bfg}:=\begin{bmatrix} \bfg^e\\ \hat{\bftau}^e\end{bmatrix},
 \end{split}
 \eeqs
 \end{widetext}
and $\tilde{\bm{\sigma}}^*$ denotes the generalized fluctuation coefficients that satisfy
 \beqs\label{sigmabar}
\tilde{\bm{\sigma}}^*\tilde{\bm{\sigma}}^{*^T}=2k_BT\begin{bmatrix}\bfQ^T\hat{\bf R}^\tt \bfQ&\bfQ^T\hat{\bf R}^\tr \\ \hat{\bf R}^{\tr^T}\bf Q& \hat{\bf R}^\rr\end{bmatrix}.
\eeqs
Suppose that the external force and torque is conservative,  meaning that  there exists a potential field $V=V(\bfxtld)$ such that the rate of work done by the external force and torque is equal to  the decrease rate   of potential energy, i.e.,
\beqs\label{solvegetaue}
\begin{split}
-\frac{\mathrm{d}}{\mathrm{d}t}V(\tilde{\bfx})&=-\bfv\cdot\nabla_\bfx V(\tilde{\bfx})-\dot{\bfTheta}\cdot \nabla_\Theta V(\tilde{\bfx})\\
&=\bfg^e\cdot{\bfv}+\hat{\bftau}^e\cdot\bfomegahat.
\end{split}
\eeqs
  From \eqref{tildex} and \eqref{solvegetaue} ,  it follows that
\beqs\label{eq: forceg}
 \bfg^e=-\nabla_\bfx V(\tilde{\bfx}); \; \qquad \;\hat{\bftau}^e=-\bfT^{T} \nabla_\Theta V(\tilde{\bfx}).
 \eeqs

As for \eqref{governingequationnew2}, we consider the PDF $P=P(\tilde{\bfx}, t)\equiv P(\bfx, \bfTheta, t)$ in the position-orientation space.  Based on \eqref{haha}, \eqref{sigmabar},  and \eqref{eq: forceg}, the Fokker-Planck equation for the PDF $P(\bfx, \bfTheta, t)$ of the stochastic process governed by \eqref{governingequationnew2} can be written as
 \begin{widetext}
 \beqs \label{eq:keyFP}
 \begin{split}
 \frac{\partial }{\partial t} P(\bfx, \bfTheta, t)=\nabla_\bfx\cdot &\Big[\Big(-\overline{\bfalpha}^\tt P(\bfx, \bfTheta, t)+\frac{1}{2}\overline{\bfbeta}^\tt\nabla_\bfx P(\bfx, \bfTheta, t)+\frac{1}{2}\overline{\bfbeta}^\tr\nabla_\bfTheta P(\bfx, \bfTheta, t)\Big)\Big]\\
+\nabla_\Theta\cdot&\Big[\Big(-\overline{\bfalpha}^\rr P(\bfx, \bfTheta, t)+ \frac{1}{2}\overline{\bfbeta}^\rr\nabla_\bfTheta P(\bfx, \bfTheta, t)+\frac{1}{2}\overline{\bfbeta}^{\tr^T}\nabla_\bfx P(\bfx, \bfTheta, t)\Big)\Big],
\end{split}
 \eeqs
 \end{widetext}
where
 \begin{widetext}
 \beqs \label{eq:alphabeta}
 \begin{split}
\begin{bmatrix}
 \overline{\bfalpha}^\tt\\
 \overline{\bfalpha}^\rr
 \end{bmatrix}=\begin{bmatrix}
\left(\bfR^\tt-\bfR^\tr\bfR^{\rr^{-1}}\bfR^{\tr^T}\right)^{-1}\nabla_\bfx V- \left(\bfR^\tt-\bfR^\tr\bfR^{\rr^{-1}}\bfR^{\tr^T}\right)^{-1}\bfR^{\tr}\bfR^{\rr^{-1}}\bfQ^{T}\bfT^{T} \nabla_\Theta V\\
 \bfT\bfQ\left(\bfR^\rr-\bfR^{\tr^T}\bfR^{\tt^{-1}}\bfR^\tr\right)^{-1}\bfQ^{T}\bfT^{T}\nabla_\Theta  V-\bfT\bfQ\left(\bfR^\rr-\bfR^{\tr^T}\bfR^{\tt^{-1}}\bfR^\tr\right)^{-1}\bfR^{\tr^T}\bfR^{\tt^{-1}} \nabla_\bfx V
 \end{bmatrix}\\
 \end{split}
 \eeqs
  \end{widetext}
and
 \begin{widetext}
 \beqs\label{key}
  \begin{split}
\begin{bmatrix}
 \overline{\bm{\beta}}^\tt
 &\overline{\bm{\beta}}^\tr\\
 \overline{\bm{\beta}}^{\tr^T}
 &\overline{\bm{\beta}}^\rr
 \end{bmatrix}&=\bm{\tilde{L}}\tilde{\bm{\sigma}}^*\tilde{\bm{\sigma}}^{*^T}\bm{\tilde{L}}^{T}=\begin{bmatrix}
 2k_BT\left(\bfR^\tt-\bfR^\tr\bfR^{\rr^{-1}}\bfR^{\tr^T}\right)^{-1}
 &-2k_BT\left(\bfR^\rr\bfR^{\tr^{-1}}\bfR^{\tt}-\bfR^{\tr^T}\right)^{-1}\bfQ^{T}\bfT^{T}\\
 -2k_BT\bfT\bfQ\left(\bfR^\tt\bfR^{\tr^{-T}}\bfR^{\rr}-\bfR^{\tr}\right)^{-1}
 &2k_BT\bfT\bfQ\left(\bfR^\rr-\bfR^{\tr^T}\bfR^{\tt^{-1}}\bfR^\tr\right)^{-1}\bfQ^{T}\bfT^{T}
 \end{bmatrix}.
  \end{split}
 \eeqs
    \end{widetext}

From \eqref{eq:keyFP},  we identify
  $ \overline{\bm{\beta}}^\tt/2$ and $\overline{\bm{\beta}}^\rr/2$  as the macroscopic translational and rotational  diffusivity, respectively. Typically,   $\bfTheta$  and  $ \bfTheta+2\pi$ refer to the same rotation. Then the crossover time scale $t_{\rm cross}$ can be estimated as
\beqs \label{timescale}
  t_{\rm cross}\sim \frac{(2\pi)^2}{|\bfbeta^\rr|}.
  \eeqs
 Moreover, it is straightforward to verify that
 \beqs \label{eq:PsTheta}
 \begin{split}
 &P^s(\bfx, \bfTheta, t)\propto \exp\Big({-\frac{V(\bfx, \bfTheta)}{k_BT}}\Big) \\
 \end{split}
 \eeqs
is a stationary solution to \eqref{eq:keyFP}, which is consistent with the classical statistical mechanics.

 We remark that  both the diffusivity tensor  $ \overline{\bm{\beta}}^\tt/2$ and $\overline{\bm{\beta}}^\rr/2$  in general depend on the current  orientation $\bfTheta$ of the particle (cf.,  \eqref{key}).
For a time scale that is much larger than the crossover time scale $t_{\rm cross}$ in \eqref{timescale}, it is anticipated that  the orientation-dependence of translational diffusivity would be averaged out. If the external force $\bfg^e\equiv 0$, i.e., $V=V(\bfTheta)$,  we may
  calculate the final effective  translational diffusivity by considering trial   solution $P=P(\tilde{\bfx}, t)$  to \eqref{eq:keyFP} of form:~\citep{Yuan2021}
  \beqs\label{Ptranslation}
P(\bfx, \bfTheta, t) =\bar{P}(\bfx, t)P^s(\bfTheta),
  \eeqs
  where $\bar{P}(\bfx, t)$ denotes the   PDF in  spatial
position $\bfx$ and
  \beqs\label{Ps}
  P^s(\bfTheta)\propto \exp\Big({-\frac{V(\bfTheta)}{k_BT}}\Big)
  \eeqs
is the  stationary  PDF in   orientation $\bfTheta$.
Inserting \eqref{Ptranslation} into \eqref{eq:keyFP} and   integrating \eqref{eq:keyFP} over the $\bfTheta$-space, we obtain  the translational diffusion equation in the long-time limit:
  \beqs \label{eq:keyFPx}
 \begin{split}
 \frac{\partial }{\partial t} \bar{P}(\bfx, t)&= \nabla_\bfx\cdot \big[\bfD^\eff \nabla_\bfx \bar{P}(\bfx, t)\big],
\end{split}
 \eeqs
where
 \beqs \label{eq:Deff general}
 \begin{split}
 \bfD^\eff  &=\int_{\SO(3)} \frac{\overline{\bm{\beta}}^\tt}{2} P^s(\bfTheta)  \\
 &=k_BT\int_{\SO(3)}\left(\bfR^\tt-\bfR^\tr\bfR^{\rr^{-1}}\bfR^{\tr^T}\right)^{-1}P^s(\bfTheta)
\end{split}
 \eeqs
 is the effective diffusivity for the long-time diffusion.

The  formula~\eqref{eq:Deff general} composes one of our main results  concerning the diffusion of general microstructured particles in an external field.     For brevity, we introduce the following  mobility tensor in the body-frame:
\beqs\label{Met}
\hat{\bfM}=\left(\hat{\bfR}^\tt-\hat{\bfR}^\tr\hat{\bfR}^{\rr^{-1}}\hat{\bfR}^{\tr^T}\right)^{-1},
\eeqs
and recall that the drag tensors $\bfRhat^{(\tt, \tr, \rr)}$, and hence the mobility tensor $\hat{\bfM}$ in the body-frame are independent of orientation parameter $\bfTheta$.
Then   \eqref{eq:Deff general} can be rewritten as
\beqs\label{eq:Deff}
\begin{split}
 \bfD^\eff  = &k_BT\int_{\SO(3)}\bfQ^T(\bfTheta)\hat{\bfM}\bfQ(\bfTheta) P^s(\bfTheta) .
 \end{split}
\eeqs
Upon inspecting ~\eqref{eq:Deff general} or \eqref{eq:Deff}, we observe    a few exact results which are listed below.

\noindent (i) Taking the trace of \eqref{eq:Deff}  we find
 \beqs \label{eq:TrD}
 \begin{split}
 \Tr(\bfD^\eff ) &= k_BT\int_{\SO(3)}  \Tr (\bfQ^T\hat{\bfM}\bfQ) P^s(\bfTheta) \\
 &=k_BT \Tr (\hat{\bfM})\int_{\SO(3)} P^s(\bfTheta) =k_BT \Tr \hat{\bfM},
\end{split}
 \eeqs
where the last equality follows from
 \beqs\label{PSdefinition}
 \int_{\SO(3)} P^s(\bfTheta) =1.
 \eeqs

\noindent (ii) If $\bfR^\tr=0$,     the translational motion is uncoupled with  the rotational  motions,  \eqref{eq:Deff general}  degenerates into
 \beqs\label{hahaha}
\bfD^\eff =k_BT\int_{\SO(3)} \bfR^{\tt^{-1}}  P^s(\bfTheta) ,
 \eeqs
 which was first  derived in Yuan et al. \citep{Yuan2021} for homogenous axisymmetric particle.

 \noindent (iii) At the absence of  external field ($V\equiv0$),    the long-time diffusivity of a free particle must be isotropic for an isotropic ambient fluid.  Then by \eqref{eq:TrD}, we immediately find the isotropic long-time diffusivity of an arbitrary microstructured free particle is given by
 \beqs\label{eq:Deff iso}
 D^\eff =\frac{1}{3} k_BT\Tr (\hat{\bfM}) .
 \eeqs

\noindent (iv) For spherical homogeneous particle of radius $r$ and at the absence of external fields,
the mobility tensor is isotropic and given by  $\hat{M}=1/{6\pi\,\eta\,r}$ ($\eta$ is the viscosity). By \eqref{eq:Deff iso} we recover the celebrated Stokes-Einstein's relation:
 \beqs\label{einstein relation}
 D^\eff =\frac{k_\text{B} T}{6\pi\,\eta\,r}.
 \eeqs

\section{Applications}\label{Sec:app}

 In this section, we present some explicit formulas for the long-time diffusivity of   heterogeneous spheroids which  account for  the effects of shape anisotropy and heterogeneities and may be regarded as  generalizations of the classical Stokes-Einstein's formula. Based on these results, we study the diffusion  of a pair of spheroids whose  relative position and orientation  are continuously relaxing from an initial non-equilibrium state to the final equilibrium state.  The additional relaxation time-scales  cause   time-dependent diffusivity of the pair. Consequently,  the  mean squared displacement ($MSD$) scales differently with respect to time   from the conventional Brownian motions because of the interplay between multiple time scales, i.e., the relaxation  time-scales,  crossover time-scale, and diffusion time-scale. We speculate that this  may be  a reasonable physical model to  understand the anomalous diffusion  observed  in, e.g., the migration of macromolecules or  cells in complex media.

\subsection{Effects of shape anisotropy and heterogeneity } \label{sec:Hetero}

Most of  existing  works  on  diffusions of microparticles assume uncoupled translational and rotational motions, i.e., the off-diagonal block $\bfR^{\tr}$ in  the drag tensor \eqref{eq:bfRblock} vanishes.  The presence of shape anisotropy and heterogeneity of the particle in general lead to the mismatch between the center of mass and the geometric centroid or hydrodynamic center of the particle and hence nonzero $\bfR^\tr$.  In this section we   consider heterogeneous   spheroidal particles  whose centers of mass deviate from their geometric centroids (or hydrodynamic centers). As will be shown shortly, the diffusivity of heterogeneous   balls could be significantly different from the conventional Stokes-Einstein's formula if the mismatch between the center of mass and geometric center is large. This fundamental fact seems to be unnoticed in the literature and may find applications in, e.g., characterizing the uniformity of microparticles or separating microparticles of different uniformity.

To proceed, we recall that the equation of motion  \eqref{eq:kinetic anisotropic000} for the particle is formulated with respect to the center of mass of the particle.  On the other hand,  the   drag matrix   is usually derived  with respect to the centroid of the particle. To make a distinction, we  denote by
 \beqs \label{eq:Upsilon0}
\begin{split}
  {\bm{\Upsilon}}:=\begin{bmatrix}{\bm{\Upsilon}}^\tt&{\bm{\Upsilon}}^\tr\\
  ({\bm{\Upsilon}}^\tr)^T &{\bm{\Upsilon}}^\rr\end{bmatrix}
 \end{split}
\eeqs
the  drag matrix with respect to the centroid of the particle and  use $\hat{{\bm{\Upsilon}}}$  (and sub-blocks $\hat{{\bm{\Upsilon}}}^{(\tt,\rr,\tr)}$) the drag matrices in the body-frame.  For spheroids, the drag matrices $\hat{{\bm{\Upsilon}}}^{(\tt,\rr,\tr)}$ depends only on the shape of the  particle and the viscosity of the ambient fluid and can be found in, e.g., the textbook of Kim and Karrila.~\citep{Kim1991Microhydrodynamics}
 The drag matrices with respect to the center of mass $\hat{\bfR}^{(\tt,\rr,\tr)}$ in terms of $\hat{{\bm{\Upsilon}}}^{(\tt,\tr,\rr)}$ can then be   obtained by a  quick free-body-diagram analysis  as (see, e.g.,  Brenner  \cite{Brenner1965Coupling,Bernal1980})
{  \beqs\label{eq: R with c}
\begin{split}
&\hat{R}^\tt_{ij}={\hat{\Upsilon}}^\tt_{ij},\\
&\hat{R}^\tr_{ij}={\hat{\Upsilon}}^\tr_{ij}-\epsilon_{isk} c_k {\hat{\Upsilon}}^{\tt}_{sj},\\
&\hat{R}^\rr_{ij}={\hat{\Upsilon}}^\rr_{ij}-\epsilon_{ipq} \epsilon_{jks} c_q c_k {\hat{\Upsilon}}^{\tt}_{ps}  +\epsilon_{jks} c_k {\hat{\Upsilon}}^{\tr}_{is}-\epsilon_{isk} c_k {\hat{\Upsilon}}^{\tr}_{sj},
\end{split}
\eeqs}
 where
 where $\bfc\in \mathbb{R}^3$ is  the vector pointing from the  centroid   to the center of mass of the particle, and $\epsilon_{iks}$  the Levi-Civita symbol.%

\begin{figure}[H]
\centering
\includegraphics[width=8.7cm]{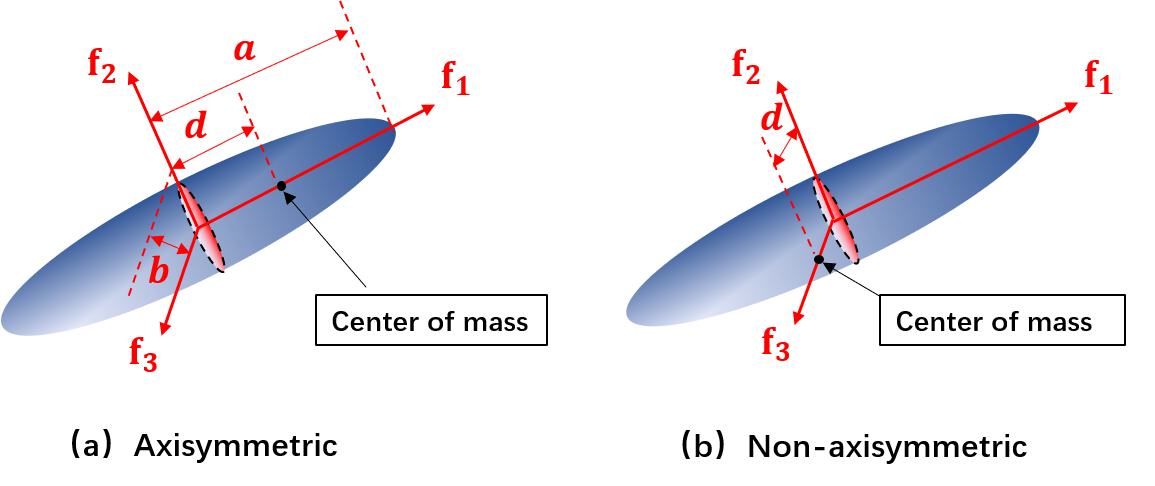}
\caption{Two typical configurations of heterogeneous micro-structured particles of prolate spheroid with axis   $\bff_1$: (a) the center of mass is on the axis of symmetry and (b) the center of mass is away from the axis of symmetry. The major semi-axis-length of the spheroidal particle is $a$, and the minor semi-axis-length of the spheroidal particle is $b$.}\label{configuration}
\end{figure}

 As illustrated in Fig.~\ref{configuration}, we will consider two  configurations of  heterogeneous   spheroidal particles.  Denote by $a$  and $b$ the two principle semi-axis-lengths of the  spheriod, and $e=a/b$ the aspect ratio. The body frame $\{\bff_i: i=1,2,3\}$ is   fixed at the centroid of the spheroidal particle with  $\bff_1$ being the axis of symmetry.
 Because of the axis-symmetry, the off-diagonal block  $\hat{\bm{\Upsilon}}^\tr $   vanishes whereas the diagonal components of $\hat{\bm{\Upsilon}}^\tt$ and $\hat{\bm{\Upsilon}}^\rr$  are given by ($\hat{\Upsilon}^\tt_{22}=\hat{\Upsilon}^\tt_{33}$,  $\hat{\Upsilon}^\rr_{22}=\hat{\Upsilon}^\rr_{33}$ and all off-diagonal components vanish by symmetry) \cite{Kim1991Microhydrodynamics} 
 \beqs\label{eq: upsilon}
 \begin{split}
&\hat{\Upsilon}^\tt_{11}=\frac{{8\pi \eta a{\left(e^2-1\right)^{\frac{3}{2}}}}}{e{\left[ {\left( 2e^2-1\right)\ln \left(e+\sqrt{e^2-1}\right) - e\sqrt{e^2-1}} \right]}},\\
&\hat{\Upsilon}^\tt_{22}= \frac{{16\pi \eta a{\left(e^2-1\right)^{\frac{3}{2}}}}}{e{\left[ {\left( 2e^2-3 \right)\ln \left(e+\sqrt{e^2-1}\right) +e\sqrt{e^2-1}} \right]}},\\
&\hat{\Upsilon}^\rr_{11}=\frac{{16\pi \eta a^3{\left(e^2-1\right)^{\frac{3}{2}}}}}{3e^3{\left[ {-\ln \left(e+\sqrt{e^2-1}\right) + e\sqrt{e^2-1}} \right]}},\\
&\hat{\Upsilon}^\rr_{22}= \frac{{16\pi \eta a^3{\left(e^2-1\right)^{\frac{3}{2}}}}(e^2+1)}{3e^3{\left[ {\left( 2e^2-1\right)\ln \left(e+\sqrt{e^2-1}\right) - e\sqrt{e^2-1}} \right]}}.
 \end{split}
 \eeqs

\subsubsection{Axisymmetric configuration} \label{sec:Aconfig}

In the first configuration illustrated in  Fig.~\ref{configuration}(a), the center of mass is on the axis of symmetry with $\bfc=d\bff_1$ for some $d\in[0,a)$.
Inserting \eqref{eq: R with c}   into  \eqref{Met} we find the mobility tensor is given by
\beqs\label{eq:Meffforf2}
\begin{split}
\hat{\bfM}=\mathrm{diag}\left(\frac{1}{{\hat{\Upsilon}}^\tt_{11}},\; \; \frac{1}{{\hat{\Upsilon}}^\tt_{22}}+\frac{d^2}{{\hat{\Upsilon}}^\rr_{22}},\;\;\frac{1}{{\hat{\Upsilon}}^\tt_{22}}+\frac{d^2}{{\hat{\Upsilon}}^\rr_{22}}\right).
\end{split}
\eeqs
Since the center of mass locates at the axis of symmetry,
it is sufficient to describe the orientation of the particle by the usual spherical coordinates  $\bfTheta=(\theta,\varphi)\in [0,\pi]\times[0,2\pi)$  for $S^2$ (unit sphere in $\rz^3$), where $\theta$ is   the angle between the symmetry axis $\bff_1$ and $\bfe_1$, and $\varphi$ is   the angle between $\bfe_2$ and the projected ray of $\bff_1$  on the $\bfe_2$-$\bfe_3$-plane.

For the diffusivity in the long-time limit, our goal is to evaluate  the integral~\eqref{eq:Deff}.  Being axis-symmetric, this integral over $\SO(3)$  reduces to an integral on $(\theta, \varphi)$ over  $S^2$. Moreover, we find the  rotation matrix $\bfQ$   as \citep{Yuan2021}
\beqs\label{Q electric2}
\begin{split}
\bfQ(\varphi ,\theta)
     =&\begin{bmatrix}
    \cos\theta    &\sin\theta  \cos\varphi & \sin\theta  \sin\varphi \\
    \sin\theta   &-\cos\theta  \cos\varphi & -\cos\theta  \sin\varphi \\
    0 & \sin\varphi & -\cos\varphi \\
    \end{bmatrix},
    \end{split}
\eeqs
and the stationary PDF $P^s(\theta,\varphi)$  in orientational space is given by (c.f., \eqref{Ps} and \eqref{PSdefinition})
\beqs\label{eq:psproperty2}
P^s(\theta,\varphi)= \frac{\sin\theta \exp\Big[-\frac{V(\theta,\varphi)}{k_BT}\Big]}{\int _0^{\pi }\!\!\!\int _0^{2 \pi } \sin\theta \exp\Big[-\frac{V(\theta,\varphi)}{k_BT}\Big]\;\mathrm{d}\theta\; \mathrm{d}\varphi}.
\eeqs
 Inserting \eqref{eq:Meffforf2}
 and \eqref{Q electric2} into \eqref{eq:Deff}, we find  the diffusivity  along the axes   $\{\bfe_1, \bfe_2, \bfe_3\}$ of the  global frame as:
 \beqs
\label{ett electricinitial}
\begin{split}
D^{\eff}_{11}=&k_BT\big[\hat{M}_{11}- (\hat{M}_{11}-\hat{M}_{22}) \omega \big]   ,\\
D^{\eff}_{22}=&k_BT\big[\hat{M}_{22}+ (\hat{M}_{11}-\hat{M}_{22})\omega'  \big]\\
D^{\eff}_{33}=&k_BT\big[\hat{M}_{22}+ (\hat{M}_{11}-\hat{M}_{22})(\omega-\omega')  \big],\\
 \end{split}
\eeqs
 where $\hat{M}_{11},\hat{M}_{22},\hat{M}_{33}$ are given by \eqref{eq:Meffforf2}, and
 \beqs \label{eq:omega}
  \begin{split}
 &\omega=\int _0^{\pi }\!\!\!\int _0^{2 \pi }\!\!\!P^s(\theta,\varphi) \sin^2\theta\;\mathrm{d}\theta\; \mathrm{d}\varphi,\\
 &\omega'=\int _0^{\pi }\!\!\!\int _0^{2 \pi }\!\!\!P^s(\theta,\varphi) \cos^2\varphi\sin^2\theta\;\mathrm{d}\theta\; \mathrm{d}\varphi,\\
  \end{split}
 \eeqs
 Once the external  potential $V(\theta,\varphi)$ is prescribed, one can simply evaluate the integrals in \eqref{eq:omega} to obtain the diffusivities in \eqref{ett electricinitial}. In general, we   expect nontrivial off-diagonal components in the diffusivity tensor.  For explicit results, we consider two special scenarios.

 \noindent (i) The external field, e.g., a strong applied  magnetic field along $\bfe_1$-direction, tends to align the spheroid axis $\bff_1$ with $\bfe_1$.
The effect of this external field can be modeled by  the external potential $V(\theta,\varphi)=E_0\sin^2\theta$. By symmetry it is easy to see all off-diagonal components of diffusivity tensor $\bfD^\eff$ vanish and $D^{\eff}_{22}=D^{\eff}_{33}$. Moreover, by   \eqref{ett electricinitial}  we find   that
 \beqs
 \label{ett electric}
\begin{split}
D^{\eff}_{11}=&k_BT\big[\hat{M}_{11}- (\hat{M}_{11}-\hat{M}_{22}) \omega \big]   ,\\
D^{\eff}_{22}=& k_BT\big[\hat{M}_{22}+ \frac{\hat{M}_{11}-\hat{M}_{22}}{2}\omega  \big],
 \end{split}
\eeqs
Let
 \beas
 \sigma^2={k_BT\over E_0}
 \eeas
 be the dimensionless parameter for the strength of alignment field. If
  $ \sigma^2   \ll 1$, i.e., the spheroid  is aligned along the field direction and weakly fluctuates, by \eqref{eq:psproperty2} the first integral in \eqref{eq:omega} is well-approximated by
 \beas
 \omega\approx{\int_0^\infty s^3 e^{-s^2/\sigma^2} ds\over \int_0^\infty s e^{-s^2/\sigma^2} ds}=\sigma^2.
 \eeas
Therefore, the diffusivities in \eqref{ett electric} are approximately given by
    \beqs
\label{ett electric1}
\begin{split}
&D^{\eff}_{11}\approx \frac{k_BT}{\Upsilonhat^\tt_{11}} [1+(\frac{\Upsilonhat^\tt_{11}}{{\Upsilonhat^\tt_{22}}}-1)\sigma^2-\frac{\Upsilonhat^\tt_{11}}{\Upsilonhat^\rr_{22}}d^2\sigma^2],\\
& D^{\eff}_{22} \approx  \frac{k_BT}{\Upsilonhat^\tt_{22}} \big[1+\frac{\Upsilonhat^\tt_{22}}{\Upsilonhat^\rr_{22}}d^2-\frac{\sigma^2}{2}(1-\frac{\Upsilonhat^\tt_{22}}{{\Upsilonhat^\tt_{11}}})+\frac{\Upsilonhat^\tt_{22}}{2\Upsilonhat^\rr_{22}}d^2\sigma^2\big],
\end{split}
\eeqs
 which may be compared with the results in  Yuan {\em et al.}\citep{Yuan2021} In particular, we notice that the  corrections in diffusivities because of the heterogeneities depend on $\sigma$. As the magnitude of the external field tends to infinity, we have $\sigma\to 0$, and hence 
 \beqs \label{eq:Daligned}
 \begin{split}
 &D^{\eff}_{11}=k_BT\Mhat_{11}=\frac{k_BT}{\Upsilonhat^\tt_{11}} ,\\
 &D^{\eff}_{22} =\frac{ k_BT(\Mhat_{22}+\Mhat_{33})}{2}=\frac{k_BT}{\Upsilonhat^\tt_{22}}(1+{\Upsilonhat^\tt_{22} d^2\over \Upsilonhat^\rr_{22}}).
 \end{split}
 \eeqs
 From the above expressions, we see that  the diffusivity along the axis-direction   is independent of the heterogeneity parameter $d$ since the spheroid is always aligned with the (strong) external field direction, i.e.,  $\bff_1\equiv\bfe_1$. In contrast, the translational motions on the transverse plane are coupled with the rotation around the axis, giving rise to $d$-dependent  diffusivity  $D^{\eff}_{22}$ in the transverse directions.

\noindent (ii) At the absence of  external field, i.e., $V(\theta,\varphi)\equiv 0$,  by directly evaluating the integrals in   \eqref{eq:psproperty2}  and  \eqref{ett electric} we find  that the diffusivity tensor is indeed isotropic,  and the diffusivity (along any direction) is given by
\beqs\label{Deffiso3}
\begin{split}
 {D}^\eff&=\frac{k_BT}{3}\frac{{\hat{\Upsilon}}^\rr_{22}(2{\hat{\Upsilon}}^\tt_{11}+{\hat{\Upsilon}}^\tt_{22})+2d^2{\hat{\Upsilon}}^\tt_{22}{\hat{\Upsilon}}^\tt_{11}}{{\hat{\Upsilon}}^\tt_{11}{\hat{\Upsilon}}^\tt_{22}{\hat{\Upsilon}}^\rr_{22}}\\
&={k_BT\over 6\pi a \eta} (\gamma_0+\gamma_d) ,
\end{split}
 \eeqs
where
\beqs \label{eq:alphad}
\begin{split}
\gamma_0&=\frac{ e \ln (e+\sqrt{e^2-1})}{ \sqrt{e^2-1} },\\
\gamma_d&={3   e^4 \sqrt{e^2-1}- 3   e^2 \left(2 e^2-1\right) \ln (e+\sqrt{e^2-1})\over 4  \left(e^2-1\right)^{3/2} \left(e^2+1\right) } {d^2\over a^2}  .
\end{split}
\eeqs
Compared with the Stokes-Einstein's formula~\eqref{einstein relation}, we recognize the dimensionless coefficients  $\gamma_0$ reflects the effect of shape anisotropy whereas the dimensionless coefficients  $\gamma_d$ signifies the importance of heterogeneity.   In particular, if the particle is spherical with  $e=1$, the diffusivity is given by ($r=a=b$)
\beqs \label{eq:Deffball}
{D}^\eff=\frac{k_\text{B} T}{6\pi\,\eta\,r}\left(1+\frac{1}{2}\frac{d^2}{r^2}\right),
\eeqs
which can be regarded as  a generalization of  the classic Stokes-Einstein's formula for heterogeneous spherical particles.

\begin{figure}[H]
\flushleft
\includegraphics[width=9.5cm]{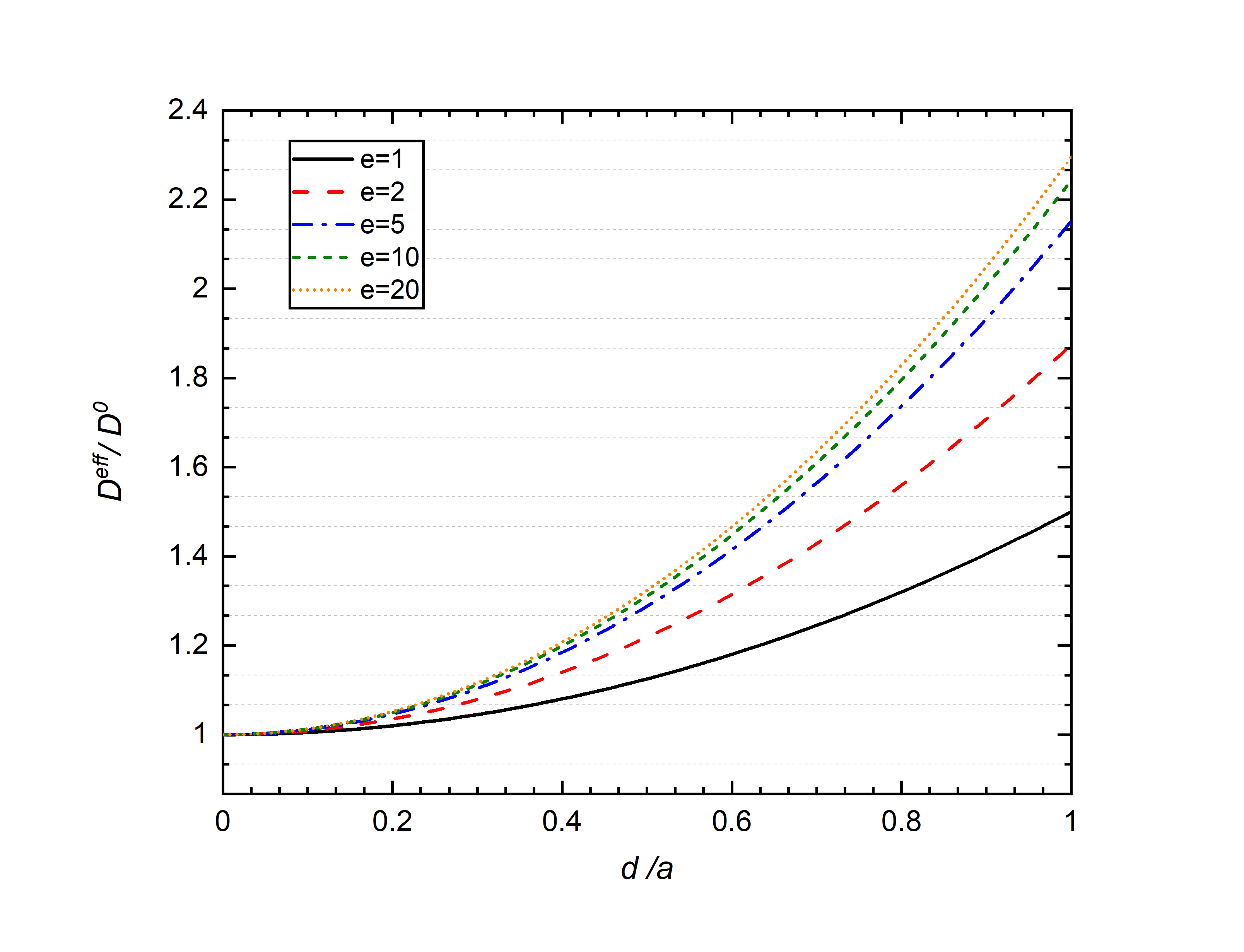}
\caption{The nondimensionalized diffusivity $D^\eff/D^\eff_0=1+\gamma_d/\gamma_0$ of   heterogeneous  spheroids as illustrated in  Fig.\protect\ref{configuration} (a)  versus the normalized distance $\frac{d}{a}$, where $d$ is the distance between the center of mass and the centroid of the spheroid,  $a$ is the   major semi-axis-length of the particle, and $e=a/b$ is the aspect ratio.}\label{inhomo2}
\end{figure}

In Fig.~\ref{inhomo2}, we   consider five  different aspect ratios  ($e=1,2,5,10,20$) at the absence of an external field and plot the normalized diffusivity $D^\eff/D^\eff_0=1+\gamma_d/\gamma_0$ versus $d/a$, where$D^\eff_0=D^\eff\big|_{d=0}$. It can be seen that the effects of  heterogeneity is more pronounced for larger aspect ratio $e$.  For a spherical particle the classic Stokes-Einstein's formula underestimates the diffusivity   by $12.5\%$  if the heterogeneity gives rise to $d/r=1/2$ (c.f., \eqref{eq:Deffball}).

\subsubsection{Non-axisymmetric configuration}

In the second configuration illustrated in Fig.~\ref{configuration}(b), the vector pointing from the centroid of the particle to the center of mass of the particle is assumed to be  $\bfc=d\bff_2$ for some $d\in[0,b)$.
Substituting \eqref{eq: R with c}  with $\bfc=d\bff_2$ into  \eqref{Met} yields
\beqs\label{eq:Meffforf1}
\begin{split}
\hat{\bfM}=\mathrm{diag}\left(\frac{1}{{\hat{\Upsilon}}^\tt_{11}}+\frac{d^2}{{\hat{\Upsilon}}^\rr_{22}},\; \; \frac{1}{{\hat{\Upsilon}}^\tt_{22}},\;\;\frac{1}{{\hat{\Upsilon}}^\tt_{22}}+\frac{d^2}{{\hat{\Upsilon}}^\rr_{11}}\right).
\end{split}
\eeqs
For this case, the center of mass deviates from  the axis of symmetry  and breaks the axis-symmetry and the integral \eqref{eq:Deff} over $\SO(3)$ can no longer be reduced to an integral over $S^2$. Nevertheless, as detailed in Appendix~\ref{Sec:AdxA} we recognize the homomorphism between rigid rotations in $\SO(3)$ and  unit quaternions $\bfq\in S^3$ (unit sphere in $\rz^4$),   and then employ spherical coordinates  $\bfTheta=(\psi,\theta,\varphi)\in U\equiv [0,\pi]\times[0,\pi]\times[0,2\pi]$ for $S^3$ to parametrize unit quaternions and  rotations. More precisely,  a   unit quaternion $\bfq=x_0+x_1\bfi+x_2\bfj+x_3\bfk $ is represented as
\beqs\label{parameterization}
\begin{split}
&x_0= \cos\psi,\\
&x_1= \sin\psi\cos\theta,\\
&x_2= \sin\psi\sin\theta\cos\varphi,\\
&x_3= \sin\psi\sin\theta\sin\varphi,
\end{split}
\eeqs
  and the associated rigid rotation $
\bfQ $ in terms of  $\bfTheta=(\psi,\theta,\varphi)$ is given by \eqref{Q electric1}.
 Moreover, the stationary PDF $P^s(\psi,\theta,\varphi)$    in orientational space is now given by (c.f., \eqref{Ps} and \eqref{PSdefinition})
 \beqs\label{eq:psproperty0}
P^s\!(\!\psi,\!\theta,\!\varphi\!)\ \propto {\sin ^2\psi \sin\!\!\theta \exp\!\!\Big[-\frac{V(\psi,\theta,\varphi)}{k_BT}\Big]} .
\eeqs
Inserting \eqref{eq:Meffforf1}
 and \eqref{Q electric1} into \eqref{eq:Deff}, we find  the diffusivity  along the axes   $\{\bfe_1, \bfe_2, \bfe_3\}$ of the  global frame as:
\beqs\label{deffpotential}
\begin{split}
D^\eff_{11}=&k_BT \Big(\lambda_{11}\hat{M}_{11} + \lambda_{12}\hat{M}_{22} + \lambda_{13}\hat{M}_{33}  \Big),\\
D^\eff_{22}=&k_BT \Big(\lambda_{21}\hat{M}_{11} + \lambda_{22}\hat{M}_{22} + \lambda_{23}\hat{M}_{33}  \Big),\\
D^\eff_{33}=&k_BT \Big(\lambda_{31}\hat{M}_{11} + \lambda_{32}\hat{M}_{22} + \lambda_{33}\hat{M}_{33}\Big),
\end{split}
\eeqs
where $\lambda_{ij}\;(i,j=1, 2,3)$,  listed in  \eqref{lambda} in  Appendix~\ref{sec:AdxB},  are dimensionless parameters that  depends on the PDF in \eqref{eq:psproperty0} and $\hat{M}_{11},\hat{M}_{22},\hat{M}_{33}$   listed in  \eqref{eq:Meffforf1}.

 Once the external  potential $V(\psi, \theta,\varphi)$ is prescribed, we can   evaluate  the integrals in  \eqref{lambda} for  $\lambda_{ij}\;(i,j=1, 2,3)$ and  obtain the diffusivities along each axis direction.  In general, we   expect nontrivial off-diagonal components in the diffusivity tensor. Below we consider two special scenarios for explicit results.

\noindent (i) The external field is a strong     field along $\bfe_1$-direction that   tends to align the spheroid axis $\bff_1$ with $\bfe_1$.   The effect of this external field can be modeled by  the external potential $V(\psi, \theta,\varphi)=E_0[1-(\bff_1\cdot\bfe_1)^2]=E_0[1-Q_{11}^2]$ $(E_0\gg k_BT)$ where the expression of $Q_{11}$ is given in $\eqref{Q electric1}_1$. Then up to the order of $O(\sigma^2)$ the parameter matrix  $[\lambda_{ij}]\;(i,j=1, 2,3)$  are given by ($\sigma^2=k_BT/E_0\ll1$)
 \beqs\label{lambdaforspheroid}
 \begin{bmatrix}
 \lambda_{ij}
 \end{bmatrix}= \half \begin{bmatrix}
2-2\sigma^2&{\sigma^2 }&  {\sigma^2 }\\
  {\sigma^2 }&  1-\sigma^2 & 1-\sigma^2 \\
 {\sigma^2 }&1-\sigma^2 &1-\sigma^2 \\
 \end{bmatrix}  +o(\sigma^2). \qquad
 \eeqs
Therefore, to the leading order approximation the diffusivities in \eqref{deffpotential} are given by
 \beqs
\label{ett electric2}
\begin{split}
&D^{\eff}_{11}{\approx} \frac{k_BT}{\Upsilonhat^\tt_{11}} [1+\frac{\Upsilonhat^\tt_{11}d^2}{\Upsilonhat^\rr_{22}}+\frac{1}{2}(\frac{2\Upsilonhat^\tt_{11}}{{\Upsilonhat^\tt_{22}}}+\frac{\Upsilonhat^\tt_{11}d^2}{\Upsilonhat^\rr_{11}}-\frac{2\Upsilonhat^\tt_{11}d^2}{\Upsilonhat^\rr_{22}}-2)\sigma^2],\\
&D^{\eff}_{22}=D^{\eff}_{33}{\approx} \frac{k_BT}{\Upsilonhat^\tt_{22}}\Big [1+\frac{\Upsilonhat^\tt_{22}d^2}{2\Upsilonhat^\rr_{11}}\\
&\quad\qquad\qquad+\frac{1}{2}(\frac{\Upsilonhat^\tt_{22}}{{\Upsilonhat^\tt_{11}}}+\frac{\Upsilonhat^\tt_{22}d^2}{\Upsilonhat^\rr_{22}}-\frac{\Upsilonhat^\tt_{22}d^2}{{\Upsilonhat^\tt_{11}}}-1)\sigma^2\Big].
\end{split}
\eeqs
As the magnitude of the external field tends to infinity, i.e., $\sigma\to 0$,  by \eqref{ett electric2} we find that
 \beas
 \begin{split}
 &D^{\eff}_{11}=k_BT\Mhat_{11}=k_BT(\frac{1}{\Upsilonhat^\tt_{11}}+ \frac{d^2}{\Upsilonhat^\rr_{22}}),\\
 &D^{\eff}_{22}= D^{\eff}_{33}= \frac{ k_BT(\Mhat_{22}+\Mhat_{33})}{2}=k_BT(\frac{1}{\Upsilonhat^\tt_{22}}+ \frac{d^2}{2\Upsilonhat^\rr_{11}}) .
 \end{split}
 \eeas
 We remark that
unlike the axisymmetric configuration discussed  earlier (c.f.,~\eqref{eq:Daligned}), the diffusivity along the axis-direction   with center of mass deviating from axis   depends on the deviation distance $d$ even if the spheroid is forced to align with the external field. This  counterintuitive effect  arises from the coupling between the rotational  and translational motions (c.f., \eqref{eq:kinetic anisotropic}), causing the fluctuation in rotations increases the fluctuation in translations and hence the diffusivity along $\bfe_1$-direction.

\noindent (ii) At the absence of  external field, i.e., $V(\psi, \theta,\varphi)\equiv 0$,   from the discussion in Appendix~\ref{Sec:AdxA} the stationary PDF $P^s(\psi,\theta,\varphi)$  can be written as
 \beqs\label{eq:psproperty1}
P^s\!(\!\psi,\!\theta,\!\varphi\!) ={1\over 2\pi^2}  \sin ^2\psi \sin\!\!\theta .
\eeqs
Upon directly evaluating the dimensionless parameters $\lambda_i$    listed in  \eqref{lambda}, we find  that the diffusivity tensor is indeed isotropic,  and the diffusivity (along any direction) is given by
 \beqs\label{Deffiso2}
 \begin{split}
 {D}^\eff=&\frac{k_BT}{3}\frac{2{\hat{\Upsilon}}^\rr_{11}{\hat{\Upsilon}}^\rr_{22}({\hat{\Upsilon}}^\tt_{11}+{\hat{\Upsilon}}^\tt_{22})+d^2{\hat{\Upsilon}}^\tt_{11}{\hat{\Upsilon}}^\tt_{22}({\hat{\Upsilon}}^\rr_{11}+{\hat{\Upsilon}}^\rr_{22})}{{\hat{\Upsilon}}^\tt_{11}{\hat{\Upsilon}}^\tt_{22}{\hat{\Upsilon}}^\rr_{11}{\hat{\Upsilon}}^\rr_{22}}\\
 =&{k_BT\over 6\pi a \eta} (\gamma_0+\gamma'_d),
 \end{split}
 \eeqs
where $\gamma_0$ is given by $\eqref{eq:alphad}_1$, and
\beqs \label{eq:alphad2}
\begin{split}
\gamma'_d&=\frac{3 e^3 \left(\left(e^2-2\right) \ln \left(\sqrt{e^2-1}+e\right)+\sqrt{e^2-1} e^3\right)}{8 \left(e^2-1\right)^{3/2} \left(e^2+1\right)} \frac{d^2}{a^2}  .
\end{split}
\eeqs
We notice that $\gamma'_d$ is   distinct  from  $\eqref{eq:alphad}_2$,  meaning that the heterogeneity along different directions has different effects on the diffusivity for shape anisotropy.

\begin{figure}[H]
\flushleft
\includegraphics[width=9.5cm]{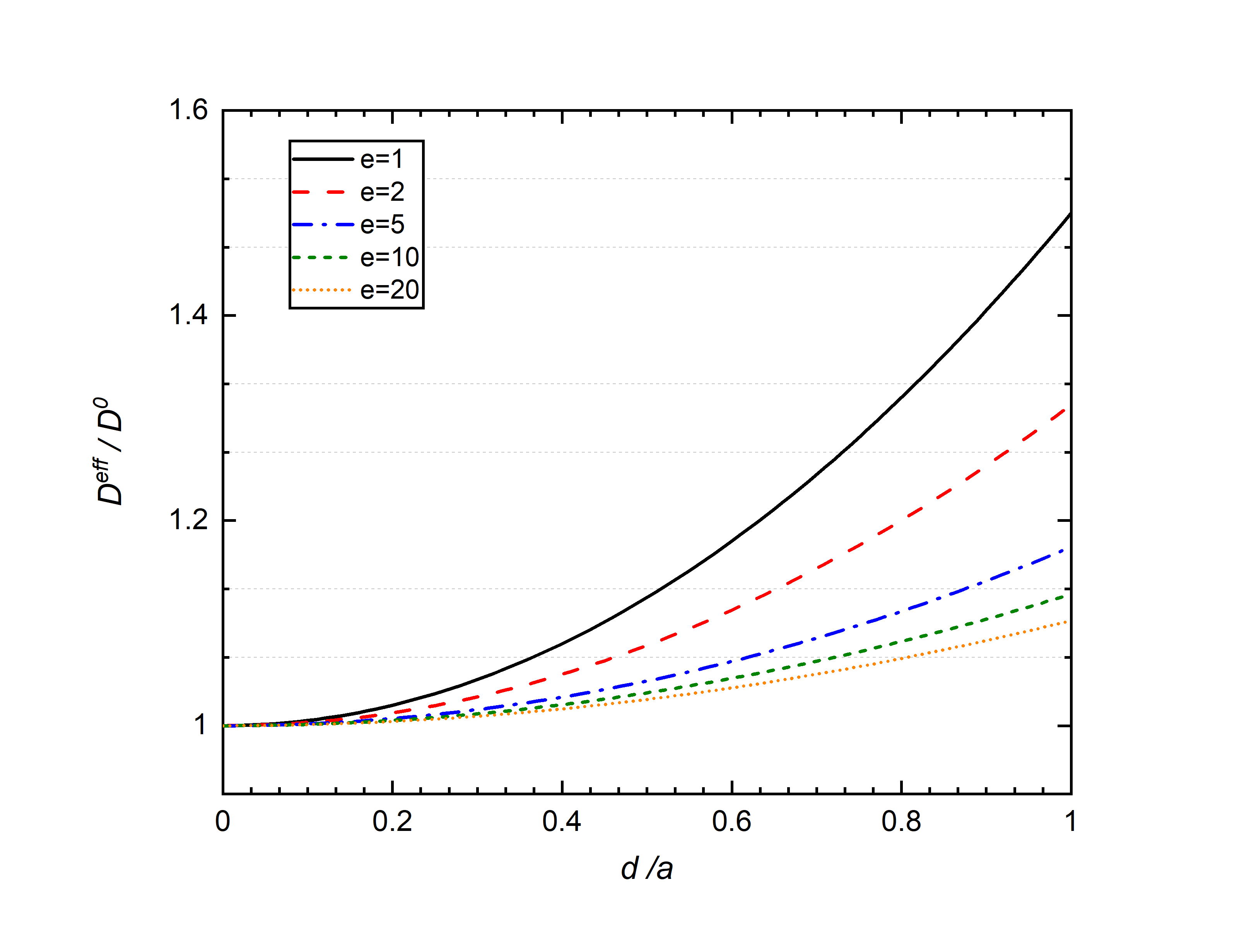}
\caption{The nondimensionalized diffusivity $D^\eff/D^\eff_0=1+\gamma'_d/\gamma_0$ of   heterogeneous  spheroids as illustrated in  Fig. \protect\ref{configuration} (b) versus the normalized distance $\frac{d}{a}$, where $d$ is the distance between the center of mass and the centroid of the spheroid,   $a $ is the   major semi-axis-length of the particle, and $e=a/b$ is the aspect ratio.}\label{inhomo1}
\end{figure}

In Fig.~\ref{inhomo1}, we  consider five  different aspect ratios  ($e=1,2,5,10,20$) at the absence of an external field and plot the normalized diffusivity $D^\eff/D^\eff_0=1+\gamma'_d/\gamma_0$ versus $d/a$.  In contrast to  Fig.~\ref{inhomo2}, we see that the impacts of  heterogeneity is more remarkable for smaller aspect ratio $e$.  In summary,  as demonstrated in  Fig.~\ref{inhomo2} and Fig.~\ref{inhomo1} the heterogeneity could significantly increase the diffusivity of particles.

\subsection{Anomalous diffusion of a pair of elastically bonded spheroids} \label{sec:Pairs}

\begin{figure}[H]
\centering
\includegraphics[width=6cm,height=5cm]{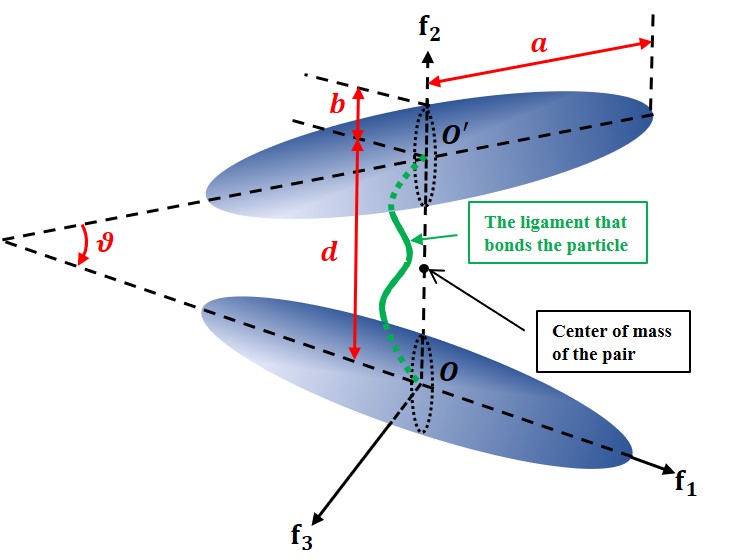}
\caption{A simple geometric structure of the pair of spheroidal micro-structured particles with only 2 degrees of freedom. The semi-axis-length parallel to the axis of symmetry of the particle is $a$, and the semi-axis-length perpendicular to the axis of symmetry of the particle is $b$. The two spheroids are bonded by some ligaments (green line) }\label{fig: Configuration pair degenerate}
\end{figure}

Based on results in  Section~\ref{sec:Hetero}, in this section  we propose a model  for anomalous diffusions by considering  a pair of spheroids in a relaxation process. As illustrated in Fig.~\ref{fig: Configuration pair degenerate},  suppose   two identical homogeneous spheroids are   bonded by some elastic ligaments.  Suppose the axes   of the two spheroids are on the same plane and form an angle $\vartheta$ and  the distance between the centroids of two spheroids is given by  $d$. We are interested in the long-time diffusivity of such a microstructured particle and how the diffusivity depends on the angle and distance $(\vartheta, d)$ and  external fields.

For simplicity, we  neglect the hydrodynamic interactions between the two spheroids in the sense that the force and torque on the centroid of each  spheroid  is given by \eqref{frictiontotal} with the drag tensor in the body frame specified by \eqref{eq: upsilon}.  By a free-body-diagram analysis,  we find the nonzero components of the blocks of the drag matrix for the pair with axis angle and separation distance $(\vartheta, d)$ can be written as
\beqs
 \begin{split}
 \hat{R}^\tt_{11}(d,\vartheta)=& \hat{\Upsilon}^\tt_{11} +\hat{\Upsilon}^\tt_{22}+(\hat{\Upsilon}^\tt_{11}-\hat{\Upsilon}^\tt_{22})\cos \vartheta,\\
 \hat{R}^\tt_{22}(d,\vartheta)=& \hat{\Upsilon}^\tt_{11} +\hat{\Upsilon}^\tt_{22}-(\hat{\Upsilon}^\tt_{11}-\hat{\Upsilon}^\tt_{22})\cos \vartheta,\\
 \hat{R}^\tt_{33}(d,\vartheta)=&2 \hat{\Upsilon}^\tt_{22},\\
  \hat{R}^\rr_{11}(d,\vartheta)=& \frac{d^2 \hat{\Upsilon}^\tt_{22}}{2}+(\hat{\Upsilon}^\rr_{11}-\hat{\Upsilon}^\rr_{22}) \cos \vartheta+\hat{\Upsilon}^\rr_{11}+\hat{\Upsilon}^\rr_{22}\\
  \hat{R}^\rr_{22}(d,\vartheta)=&(\hat{\Upsilon}^\rr_{22}-\hat{\Upsilon}^\rr_{11}) \cos \vartheta+\hat{\Upsilon}^\rr_{11}+\hat{\Upsilon}^\rr_{22},\\
 \hat{R}^\rr_{33}(d,\vartheta)=&\frac{1}{4} \left(d^2 (\hat{\Upsilon}^\tt_{11}-\hat{\Upsilon}^\tt_{22}) \cos \vartheta+d^2 (\hat{\Upsilon}^\tt_{11}+\hat{\Upsilon}^\tt_{22})+8 \hat{\Upsilon}^\rr_{22}\right)\\
\hat{R}^\tr_{32}(d,\vartheta)=&\frac{1}{2} d (\hat{\Upsilon}^\tt_{11}-\hat{\Upsilon}^\tt_{22}) \sin \vartheta,\\
 \end{split}
\eeqs
  where $\hat{{\Upsilon}}^\tt_{11}, \hat{{\Upsilon}}^\tt_{22}$ and $\hat{{\Upsilon}}^\rr_{11}, \hat{{\Upsilon}}^\rr_{22}$, given by  \eqref{eq: upsilon},  are the components of the diagonal blocks of the drag matrix for a single spheroid. 

We first consider the diffusion of  the pair with  fixed $(d,\vartheta)$.  At the absence of external fields,  the stationary PDF $P^s(\psi,\theta,\varphi)$  is given by \eqref{eq:psproperty1}.  Upon repeating the procedure from \eqref{eq:Meffforf1} to \eqref{Deffiso2},
    we find that the diffusivity tensor is indeed isotropic,  and the diffusivity (along any direction) is given by
\beqs\label{deffforthepair}
\begin{split}
&D^\eff (\vartheta, d)=\!\!\!\frac{k_BT}{3}\!\Big[\frac{1}{2\hat{\Upsilon}^\tt_{22}}+\frac{1}{ 2\hat{\Upsilon}^\tt_{11} \cos^2 \frac{\vartheta}{2} +2\hat{\Upsilon}^\tt_{22} \sin^2 \frac{\vartheta}{2} }+\\
&\quad+\frac{4\hat{\Upsilon}^\rr_{22}+d^2( \hat{\Upsilon}^\tt_{11} \cos^2 \frac{\vartheta}{2} +\hat{\Upsilon}^\tt_{22} \sin^2 \frac{\vartheta}{2})}{8\hat{\Upsilon}^\rr_{22}(\hat{\Upsilon}^\tt_{11} \sin^2 \frac{\vartheta}{2} +\hat{\Upsilon}^\tt_{22} \cos^2 \frac{\vartheta}{2})+2d^2\hat{\Upsilon}^\tt_{11}\hat{\Upsilon}^\tt_{22}}\Big].
\end{split}
\eeqs
{As $d\to +\infty$, i.e., the two spheroids are far apart, we have
\beqs\label{converge}
\begin{split}
  D^\eff \to
&\frac{k_BT}{3}\!\Big[\frac{1}{ 2\hat{\Upsilon}^\tt_{22}}+\frac{(\hat{\Upsilon}^\tt_{11}-\hat{\Upsilon}^\tt_{22}) \cos \vartheta+\hat{\Upsilon}^\tt_{11}+\hat{\Upsilon}^\tt_{22}}{4 \hat{\Upsilon}^\tt_{11} \hat{\Upsilon}^\tt_{22}}\\
&+\frac{1}{(\hat{\Upsilon}^\tt_{11}-\hat{\Upsilon}^\tt_{22}) \cos \vartheta+\hat{\Upsilon}^\tt_{11}+\hat{\Upsilon}^\tt_{22}}\Big].
\end{split}
\eeqs
}

{In particular, if $\vartheta=0$ or $\vartheta=\pi$, i.e., the two spheroids are aligned, we have
\beqs\label{deffforthepairal}
\begin{split}
 D^\eff =  \frac{k_BT}{6}\Big[ \frac{ 1}{ \hat{\Upsilon}^\tt_{11}   }+ \frac{ 2}{  \hat{\Upsilon}^\tt_{22} } \Big],
\end{split}
\eeqs
which can also be derived from \eqref{hahaha} since $\bfR^\tr={\bf 0}$. }

If, in particular,  the spheroid is a ball such that $\hat{\Upsilon}^\tt_{22}=\hat{\Upsilon}^\tt_{11}$,
\beas
 D^\eff  
 = \frac{ k_BT}{2 \hat{\Upsilon}^\tt_{11}} ,
 \eeas
 which is precisely half of the diffusivity of a single spherical particle.  

\begin{figure}[H]
\centering
\includegraphics[width=9.3cm]{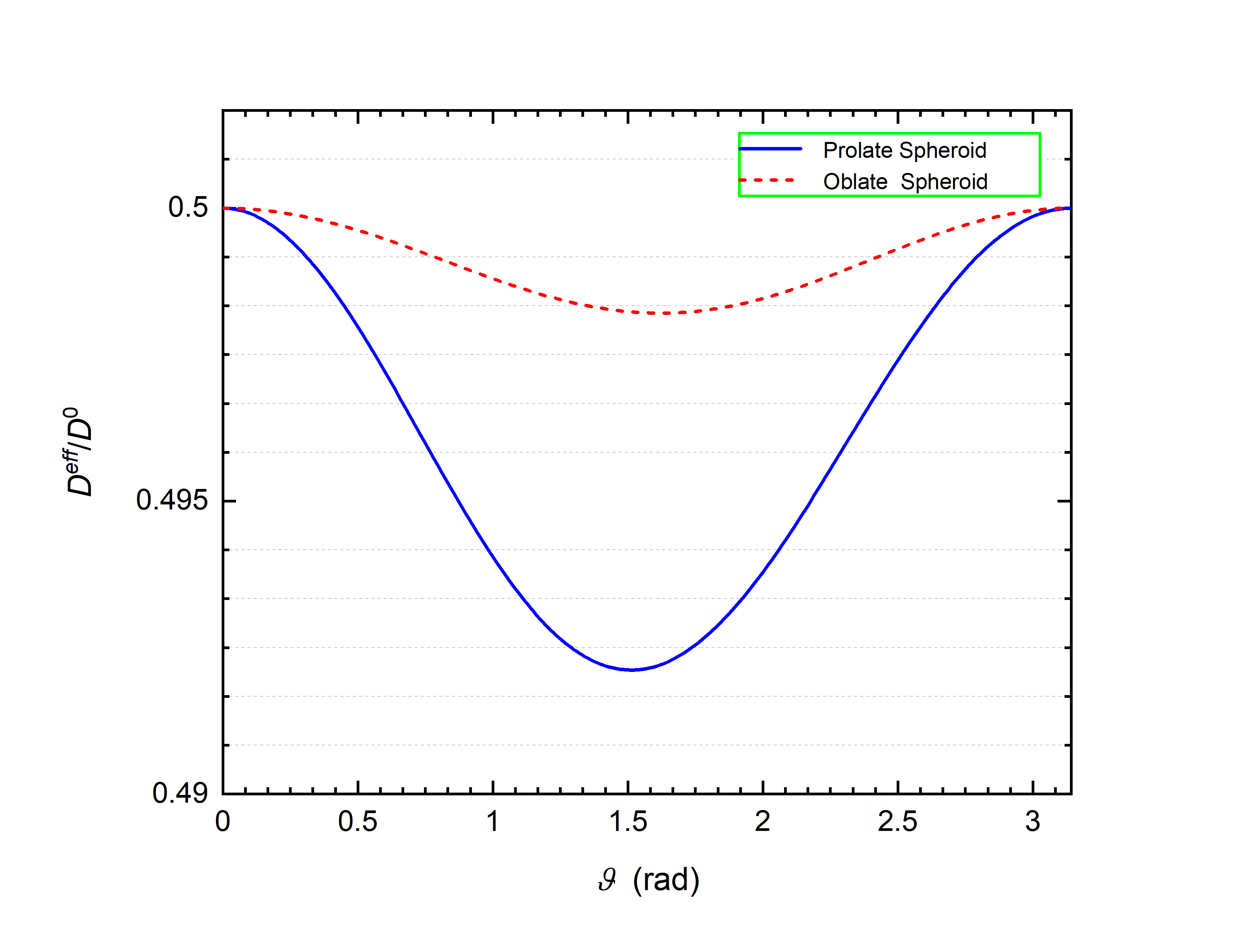}
\caption{
The nondimensionalized  long-time diffusivity $\frac{{D}^\eff}{D^0}$ of the pair of spheroids (as illustrated in Fig.\protect\ref{fig: Configuration pair degenerate})  versus   the angle $\vartheta$ between two axes of  spheroids. The major and minor semi-axis-length of spheroid are    $1 \mu m$ and $0.1 \mu m$  for prolate spheroids (blue), respectively. The  oblate spheroids   (red) have the same volume and minor-axis length. The distance $d$ between the centroids of the two particles is fixed at $2\mu m$. 
}\label{pairtheta}
\end{figure}

\begin{figure}[H]
\centering
\includegraphics[width=9.3cm]{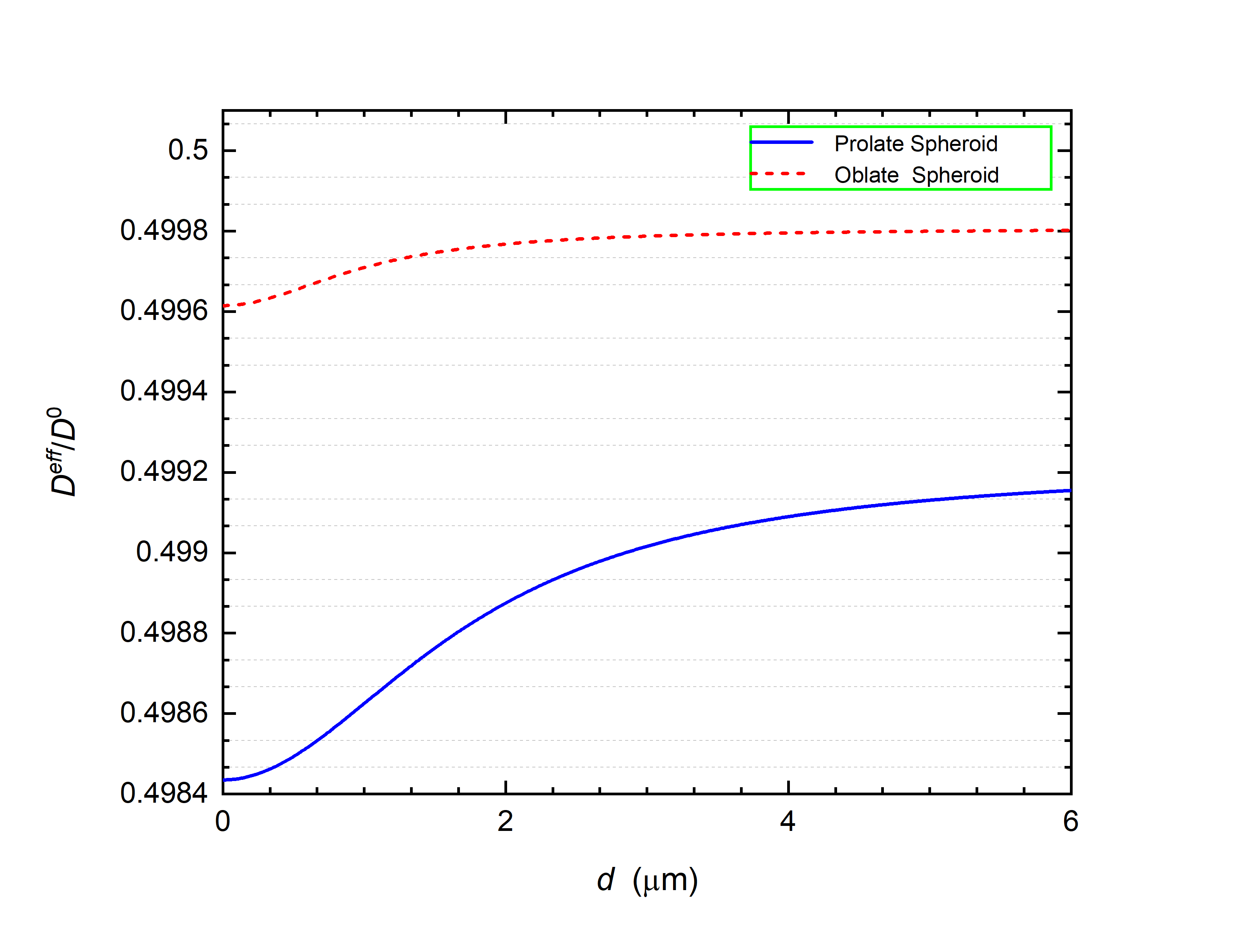}
\caption{
 The nondimensionalized  long-time diffusivity $\frac{{D}^\eff}{D^0}$ of the pair of spheroids (as illustrated in Fig.\protect\ref{fig: Configuration pair degenerate})  versus the distance $d$ between two spheroids. The major and minor semi-axis-length of spheroid are    $1 \mu m$ and $0.1 \mu m$  for prolate spheroids (blue), respectively. The  oblate spheroids   (red) have the same volume and minor-axis length.   The angle $\vartheta$ between the two axes of symmetry   is fixed at $\frac{\pi}{2}$.
}\label{paird}
\end{figure}

To see the influence of structural parameters $(\vartheta, d)$ of the pair  on the long-time diffusivity, we consider two cases: a pair of  prolates  with major and minor semi-axis-length $1 \mu m$ and $0.1 \mu m$, respectively, and  a pair of   oblates  with the same volume and minor axis-length.  In Fig.~\ref{pairtheta} we plot   the normalized diffusivity $D^\eff/D^\eff_0$ versus $\vartheta$ for { $d=2\mu m$}, where $D^0=\frac{k_BT}{3}\Tr \;[\hat{{\bm{\Upsilon}}}^{\tt^-1}]$ denotes the diffusivity of a single spheroid. Meanwhile,   the normalized diffusivity $D^\eff/D^\eff_0$ is plotted against $d$ in Fig.~\ref{paird}   for $\vartheta=\pi/2$.

From Fig.~\ref{pairtheta} we see that the diffusivity  ${D}^\eff$  of the pair is  unsurprisingly  lower than that of a single particle since the size of the pair is larger (c.f., the Stokes-Einstein relation~\eqref{einstein relation}).  On the other hand,  there exist   some optimal angles $\vartheta\in [0,\pi]$ for which the  diffusivity ${D}^\eff$ of the pairs  is  either minimized or maximized.
From Fig.~\ref{paird}  we see that the long-time diffusivity monotonically increases with $d$  until the curves flatten out, as is expected by \eqref{deffforthepair} and \eqref{converge}.

 Based on the explicit formula~\eqref{deffforthepair}, we next consider the model of a pair of spheroids  whose relative angle and distance depends on time.  Suppose the  relative angle and distance  pair are initially  given by
 $(\vartheta_0, d_0)$ and the elastic ligament is not fully relaxed. Because of the elastic energy in the ligament, we anticipate the relaxation of angle and distance can be characterized by two relaxation time scales (${\tau^*_1}, \tau^\ast_2$) in the sense that the time-dependent  angle and distance  are given by
\beqs\label{springd}
\begin{split}
\vartheta(t)&=\vartheta_f+(\vartheta_0- \vartheta_f)e^{-\frac{t}{\tau^*_1}},\\
d(t)&=\mathrm{d}_f+(\mathrm{d}_0-\mathrm{d}_f)e^{-\frac{t}{\tau^*_2}},\\
\end{split}
\eeqs
where  $(\vartheta_f, d_f)$ denote  the angle and distance between the pair in the final equilibrium state.

In  typical experimental measurements, the diffusivity or anomalous diffusion is characterized by the Mean Square Displacement ($MSD$) defined by \citep{Metzler2014}
\beqs\label{MSD0}
MSD(t):=\frac{1}{t_m}\int_0^{t_m} |\bfx(t'+t)-\bfx(t')|^2\;\mathrm{d}t',
\eeqs
where $\bfx(t)$ is the position of the center of mass  of the pair, and $t_m$ is the total measure time. For normal diffusions with a single time scale, e.g., diffusion of a homogeneous rigid spherical particle,   the scaling of  $MSD$ in \eqref{MSD0}  with respect to time  satisfies
\beqs\label{MSD00}
{MSD(t) \over  2t}\to {D} \qquad \aas\;t\to +\infty ,
\eeqs
where $D$ is the macroscopic diffusivity.  Now  we consider the scaling of $MSD$ with respect to time for the  pair of spheroids that relax from an initial non-equilibrium state. From the prior discussions, this process involves at least four time scales: cross-over time scale $t_{cross}$ (c.f., \eqref{timescale}), the two relaxation time scales $\tau^*_1, \tau^*_2$ for evolution of the relative angle and distance $(\vartheta, d)$ between the pair, and the translation diffusion time-scale of the pair. Therefore, the $MSD$ of the pair should exhibit   much more complicated scaling behaviors with respect to $t$.

Nevertheless, if there is separation of   time scales in the sense that
\beas
t\gg \tau^*_1, \tau^*_2\gg t_{cross},
\eeas
  by \eqref{MSD00} we expect that $MSD$ should approximately behave as
\beqs\label{MSD}
MSD(t) \approx  2t {D}^\eff(\vartheta(t), d(t)),
\eeqs
where the diffusivity ${D}^\eff(\vartheta(t), d(t))$ is given by \eqref{deffforthepair} and $(\vartheta(t), d(t))$ is specified by  \eqref{springd}.

\begin{figure}[H]
\centering
\includegraphics[width=9.3cm]{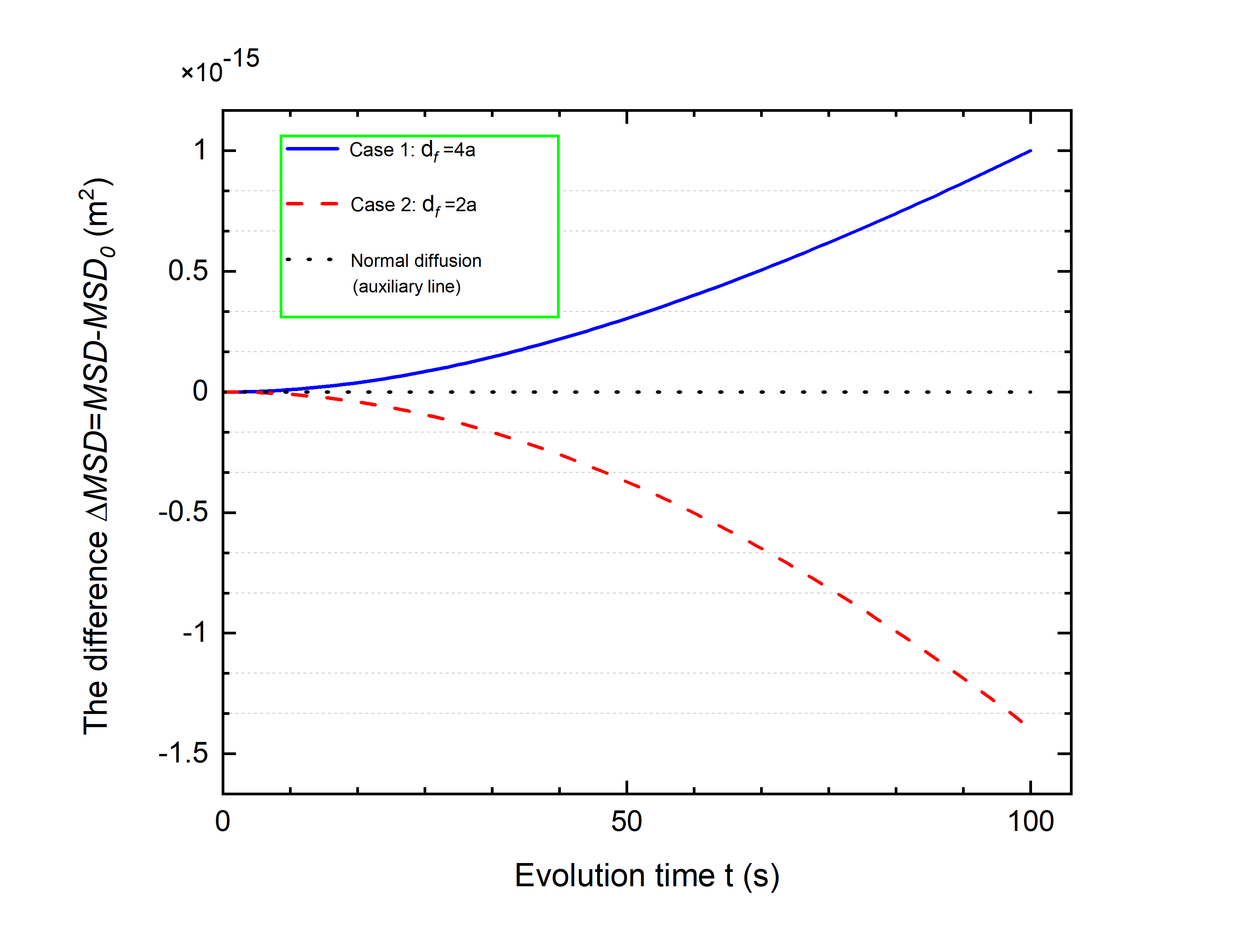}
\caption{
   {The difference $\Delta MSD=MSD-MSD_0$ between the MSDs}  of the center of mass of the pair  against time $t$ for two cases. The blue line refers to case 1  for which the final equilibrium  distance $d_f=4a$, whereas the red line refers to case 2 with the final equilibrium  distance $d_f=2a$. The distance between the pair in the initial non-equilibrium is $d_0=3a$. The major and minor semi-axis-lengths of the particle are $1 \mu m$ and {$0.1 \mu m$}, respectively.
The angle $\vartheta$  between the axes of symmetry of the pair of the particles is fixed at $\frac{\pi}{2}$. The relaxing time-scale is assumed to be $\tau^*_2=100s$. The ambient fluid is the mineral oil with viscosity $\eta=12.7\times 10^{-3}\;N\cdot s/ m^2$.
}\label{superdiffusion}
\end{figure}

{For comparison, let
\beqs\label{MSD0}
MSD_0(t) =2t {D}^\eff(\vartheta_0, d_0),
\eeqs
denote the Mean Square Displacement for the normal diffusion of the pair with $\vartheta(t)\equiv\vartheta_0$ and $d(t)\equiv d_0$.
Then, in Fig.~\ref{superdiffusion}, we plot the difference $\Delta MSD=MSD-MSD_0$ between the MSDs of the center of mass of the pair against $t$ for two cases.}
Specifically,  the distance  $d_f$ between the pair in the final equilibrium state is set to be $4a$ for case~1, whereas the distance  $d_f$ between the pair in the final equilibrium state is set to be $2a$ for case~2. The initial distance  $d_0$ between the pair in the initial non-equilibrium state is set to be $3a$ for both cases.

From Fig.~\ref{superdiffusion}   we observe  two different types of scalings fo MSD  versus $t$, which may be interpreted as   ``superdiffusion'' and ``subdiffusion'', respectively.    From this viewpoint,  we expect  physical   models like ours  could shed light on many anomalous diffusions observed in biological systems and complex media.

\section{Conclusion remarks} \label{sec:con}

In summary, we have conducted a systematic analysis on the long-time diffusion of microstructured particles of arbitrary shape and heterogeneity in a Newtonian fluid. The microscopic  Brownian dynamics in position-orientation space is coarse-grained into  a Fokker-Planck equation governing the evolution of the PDF on  $\rz^3\times \SO(3)$. By analyzing the Fokker-Planck equation, we identify a  formula for the long-time  diffusivity of the microstructured particle in an alignment field (c.f.,~\eqref{eq:Deff general} or \eqref{eq:Deff}).   Applied to heterogeneous spheroids, we discover generalizations of the classical Stokes-Einstein relation which assert that the diffusivity of a rigid heterogeneous  particle depends on the deviation of the center of mass from the geometric centroid (c.f.,~\eqref{Deffiso3}, \eqref{eq:Deffball} and \eqref{Deffiso2}). We have also addressed the effects of an external alignment field  and achieved leading-order corrections on the diffusivities for large alignment fields (c.f.,~\eqref{ett electric1} and \eqref{ett electric2}). Based on these results, we consider diffusion of a pair of spheroids bonded by elastic ligaments. If the pair are initially in a non-equilibrium state, the  Brownian motion  superimposed with the relaxation process  could be characterized as apparent ``superdiffusion'' or ``subdiffusion'', entailing a mechanistic perspective on anomalous diffusions.  We believe that our method and results  lay a solid foundation for understanding   diffusions in complex media and may inspire new   applications of controlled diffusions in biophysics and materials science.

\appendix

\section{Evaluation of integrals over $\SO(3)$ } \label{Sec:AdxA}
To find the effective long-time diffusivities of microstructured particles, we have to address the technical problem~\eqref{eq:Deff} of  evaluating integrals   over $\SO(3)$.   A  fundamental theorem in the theory of Lie group asserts that there exists a unique measure on every compact Lie group, namely, the Haar measure $d\mu_\bfQ$, such that  the integral \citep{Lie}
\beqs\label{integralSO30}
I[f]=\int_{SO(3)} f(\bfQ)\mathrm{d}\mu_\bfQ
\eeqs
is normalized (i.e., $I[1]=1$) and (left-)invariant:
\beas
  I[f(\bfQ)]=I[f(\bfQ'\bfQ)]\quad\forall\;\bfQ'\in \SO(3).
\eeas
The integral \eqref{eq:Deff} should be interpreted as in \eqref{integralSO30} for the Haar measure.

In practice, we need a   parametrization of $\SO(3)$ to carry out the integral~\eqref{integralSO30} explicitly. Here we employ the quaternion representation of $\SO(3)$.
A quaternion $\bfq$ can be written   as
\beqs\label{eq: quaternion definition}
\bfq=x_0+x_1\bfi+x_2\bfj+x_3\bfk,
\eeqs
where $\bfi,\bfj,\bfk$ are three (linearly independent) symbols. Equipped with  the regular scalar product and vector addition, the collection of quaternions form a four-dimensional vector space.  The norm of a   quaternion  is defined as
\beqs\label{unit}
|\bfq|^2=x_0^2+x_1^2+x_2^2+x_3^2.
\eeqs

Non-commutative multiplications between quaternions are defined by requiring the following relations:
\beqs\label{multiplication}
\begin{split}
&\bfi^2=\bfj^2=\bfk^2=\bfi\bfj\bfk=-1.
\end{split}
\eeqs
From \eqref{multiplication} it is straightforward to  verify
\beqs
\begin{split}
&\bfi\bfj=-\bfj\bfi=\bfk;\;\;\bfj\bfk= -\bfk\bfj=\bfi;\;\;\bfk\bfi= -\bfi\bfk=\bfj.
\end{split}
\eeqs
 Therefore, the multiplication of two   quaternions $\bfq_1=(u_1,\bfv_1)$ and $\bfq_2=(u_1,\bfv_2)$  ($u_1, u_2\in\rz$ and $\bfv_1, \bfv_2\in\rz^{3}$)  can be written  as
\beqs\label{quaternionmultiplication}
\bfq_1\bfq_2=(u_1u_2-\bfv_1\cdot\bfv_2)+ u_1\bfv_2+u_2\bfv_1+\bfv_1\times\bfv_2,
\eeqs
where `$\cdot$' and `$\times$' denote  the familiar dot product and cross product, respectively. Define $\overline{\bfq}=(x_0,-\bfx)$ as the conjugate of $\bfq=(x_0,\bfx)$. From  \eqref{quaternionmultiplication} it is straightforward to verify that  \beqs
|\bfq|^2=\bfq \bfqbar=x_0x_0+\bfx\cdot \bfx\quad\aand\quad
\overline{\bfq_1\bfq_2}=\bfqbar_2\bfqbar_1.
\eeqs
  Therefore, the inverse $\bfq^{-1}$ of a {\em unit} quaternion $\bfq\in S^3$, i.e., the quaternion such that $\bfq\bfq^{-1}=\bfq^{-1}\bfq=1$, is precisely the conjugate $\overline{\bfq}$. In addition, for two {\em unit} quaternions $\bfq_1, \bfq_2$, by~\eqref{quaternionmultiplication}  we have
\beas
|\bfq_1\bfq_2|^2=\bfq_1\bfq_2 \overline{\bfq_1\bfq_2}=|\bfq_2|^2|\bfq_1|^2=1.
\eeas
In conclusion, the collection of unit quaternions equipped with multiplication~\eqref{quaternionmultiplication} forms a continuous compact group.


Next, we construct a homomorphism   $h:S^3\to \SO(3)$ from the  group of unit quaternions to the group of rigid rotations. Let  $\bfQ=h(\bfq)$ and $\tilde{\bfp}=(0,\bfp)$  be the quaternion associated with a   vector $\bfp \in \rz^3$. For a given unit quaternion $\bfq\in S^3$, the map $\bfq\mapsto \bfQ=h(\bfq)$ is defined by requiring that for any $\bfp\in \rz^3$,
\beqs \label{eq:qtoQ}
\begin{split}
\bfp'=&\bfq\tilde{\bfp}\bfq^{-1}\quad\mathrm{quaternion\;multiplication}\\
=&\bfQ\bfp\;\quad\quad\mathrm{matrix\;product}.\\
\end{split}
\eeqs
For a unit quaternion  $\bfq$  given by  \eqref{eq: quaternion definition},   the above equation implies
  \beqs\label{correspondence}
\begin{split}
\bfp'\!&=\!(\bfp\cdot\bfx)\bfx\!+\!x_0(x_0\bfp-\bfp\!\times\!\bfx)\!+\!\bfx\!\times\!(x_0\bfp-\bfp\!\times\!\bfx)\\
&=\bfQ\bfp \qquad \forall\;\bfp\in \rz^3,
\end{split}
\eeqs
 and hence,
 \beqs\label{Qq}
 \bfQ=(x_0^2-|\bfx|^2)\bfI+2\bfx\otimes \bfx+2x_0 \bfW_\bfx,
 \eeqs
 where $\bfW_\bfx$ is the skew-symmetric matrix such that $\bfW_\bfx\bfp=\bfx\times \bfp$ for any $\bfp\in \rz^3$.
  Conversely,  for any $\bfQ\in \SO(3)$ we solve \eqref{Qq} for the unit quaternion $\bfq$ and find that
\beas
\begin{split}
&x_0=\frac{\sqrt{1+\Tr\bfQ}}{2},\\
&x_1=\sgn(Q_{32}-Q_{23})|\frac{1}{2}\sqrt{1+\Tr\bfQ-2Q_{22}-2Q_{33}}|,\\
&x_2=\sgn(Q_{13}-Q_{31})|\frac{1}{2}\sqrt{1+\Tr\bfQ-2Q_{11}-2Q_{33}}|,\\
&x_3=\sgn(Q_{21}-Q_{12})|\frac{1}{2}\sqrt{1+\Tr\bfQ-2Q_{11}-2Q_{22}}|,\\
\end{split}
\eeas
where `$\sgn$' denotes the sign function.

From \eqref{eq:qtoQ},  we find that  for any two unit quarternions $\bfq_i\in S^3$ and $\bfQ_i=h(\bfq_i)$ ($i=1,2$),
\beqs\label{coincidence2}
\begin{split}
(\bfq_2\bfq_1)\tilde{\bfp}(\bfq_2\bfq_1)^{-1} = \bfq_2(\bfq_1\tilde{\bfp}\bfq_1^{{-1}})\bfq_2^{{-1}}\\
=\bfq_2(0,\bfQ_1\bfp)\bfq_2^{{-1}}=(0,\bfQ_2\bfQ_1\bfp),
\end{split}
\eeqs
which means
\beqs  \label{eq:hbfq}
h(\bfq_2\bfq_1)=h(\bfq_2)h(\bfq_1).
\eeqs
That is, the map $\bfq\mapsto \bfQ=h(\bfq)$ defined by \eqref{eq:qtoQ}  is a homomorphism.  Therefore, the integral~\eqref{integralSO30} over $\SO(3)$ can be rewritten as integrals over $S^3$:
  \beqs\label{integralSO3} I[f]=\int_{SO(3)} f(\bfQ)\mathrm{d}\mu_\bfQ=\int_{S^3} f(\bfQ(\bfq))\mathrm{d}\mu_\bfq,
  \eeqs  where $\mu_\bfq$ represents the Haar measure on the group of unit quaternions. On $S^3$,  for any fixed unit quaternion $\bfalpha\in S^3$ and $\bfq'=\bfalpha \bfq$  we have  \beas
  d\bfq'\cdot d\bfq'=(\bfalpha  d\bfq) \overline{\bfalpha d\bfq}=d\bfq\cdot d\bfq(\alpha\overline{\alpha})=d\bfq\cdot d\bfq.
  \eeas
 That is,    the usual (normalized) Lebesgue measure
 is invariant, and hence the   Haar measure.

 Finally, we parametrize $S^3$ by the standard spherical coordinates  $\bfTheta=(\psi,\theta,\varphi)\in U\equiv [0,\pi]\times[0,\pi]\times[0,2\pi]$  so that a unit quaternion   $\bfq\in S^3$ can be represented as \eqref{eq: quaternion definition} with
 \beqs\label{parameterizationappendix}
\begin{split}
&x_0= \cos\psi,\\
&x_1= \sin\psi\cos\theta,\\
&x_2= \sin\psi\sin\theta\cos\varphi,\\
&x_3= \sin\psi\sin\theta\sin\varphi.
\end{split}
\eeqs
Then the (normalized) Lebesgue measure  on $S^3$ can be written as
\beas
d\mu_\bfq\!\propto |\det (\bfg^T\bfg)|^{1/2}\mathrm{d}\psi \mathrm{d}\theta \mathrm{d}\varphi=\sin^2\psi\sin\theta\mathrm{d}\psi  \mathrm{d}\theta \mathrm{d}\varphi,
\eeas
where $\bfg$ is the $4\times 3$ Jacobian matrix:
\beas
\bfg={\partial(x_0, x_1, x_2, x_3)\over \partial (\psi, \theta, \varphi)}.
\eeas
 By direct integration we find the volume of the hypersurface $S^3$ is given by
\beqs
V_0=\int_U\sin ^2\psi \sin\theta\mathrm{d}\psi\mathrm{d}\theta\mathrm{d}\varphi=2\pi^2.
\eeqs
Consequently, by \eqref{integralSO3} we conclude that  the integral \eqref{integralSO30} can be calculated by
\beqs
I[f]\!\!={1\over 2\pi^2}\int_U f(\bfQ(\psi,\theta,\varphi))\sin ^2\psi \sin\theta\mathrm{d}\psi\mathrm{d}\theta\mathrm{d}\varphi
\eeqs
where $\bfQ(\psi,\theta,\varphi)$, by \eqref{Qq} and \eqref{parameterizationappendix},   is given  by
\beqs\label{Q electric1}
\begin{split}
Q_{11}=&\frac{2 \cos 2 \psi-\cos 2 (\psi-\theta)+2 \cos2\theta-\cos2(\psi+\theta)+2}{4};\\
Q_{21}= &2 \sin \psi \sin \theta (\cos \theta \cos \varphi \sin \psi+\cos \psi \sin \varphi); \\
Q_{31}=&2 \sin \psi \sin \theta (\cos \theta \sin \psi \sin \varphi-\cos \psi \cos \varphi); \\
Q_{12}=&2 \sin \psi \sin \theta (\cos \theta \cos \varphi \sin \psi-\cos \psi \sin \varphi);\\
Q_{22}= &\cos ^2\psi+\sin ^2\psi \left(\cos 2\varphi \sin ^2\theta-\cos ^2\theta\right); \\
Q_{32}=&2 \sin \psi \left(\cos \varphi \sin \psi \sin \varphi \sin ^2\theta+\cos \psi \cos \theta\right); \\
Q_{13}=&2 \sin \psi \sin \theta (\cos \psi \cos \varphi+\cos \theta \sin \psi \sin \varphi); \\
Q_{23}= &2 \sin \psi \left(\cos \varphi \sin \psi \sin ^2\theta \sin \varphi-\cos \psi \cos \theta\right); \\
Q_{33}=&\cos ^2\psi-\sin ^2\psi \left(\cos ^2\theta+\cos 2\varphi \sin ^2\theta\right).
\end{split}
\eeqs

 \section{The expressions of $\lambda_{ij}$ in \eqref{deffpotential}} \label{sec:AdxB}

Inserting \eqref{Q electric1} into \eqref{eq:Deff}, we can write the diagonal components of the  diffusivity tensor~\eqref{eq:Deff} as   \eqref{deffpotential}. Recall that the mobility tensor $\hat{\bfM}$ in~\eqref{eq:Deff} is given by  \eqref{eq:Meffforf1} and the rigid rotation matrix $\bfQ(\Theta)$ in~\eqref{eq:Deff} is listed in \eqref{Q electric1}. Straightforward algebraic calculations yield the   expressions of  $\lambda_{ij}\;(i,j=1, 2,3)$ in terms of $(\psi, \theta, \varphi)$ as follows:

\beqs\label{lambda}
\begin{split}
 \lambda_{ij}=\!\!\!\int_U  Q_{ij}^2(\psi, \theta, \varphi)P^s(\psi,\theta,\varphi)\mathrm{d}\psi\mathrm{d}\theta \mathrm{d}\varphi.
 \end{split}
\eeqs
  For heterogeneous  spheroids as illustrated in  Fig. \ref{configuration} (b), if   the external field  is large in the sense that $\sigma^2=k_BT/E_0\ll 1$, then the PDF in \eqref{eq:psproperty0}  can be approximated by
  \beqs\label{eq:psproperty1}
  \begin{split}
P^s\!(\!\psi,\!\theta,\!\varphi\!) & \propto {\sin ^2\psi \sin\!\!\theta \exp\!\!\Big[-\frac{ E_0(1-Q_{11}^2)}{k_BT}\Big]} \int_0^\pi \;d\theta \\
&\approx  {\sin ^2\psi \sin\!\!\theta \exp\!\!\Big[-\frac{   4 \theta^2 (\sin\psi)^2  }{\sigma^2}\Big]} \int_0^\infty \;d\theta .
\end{split}
\eeqs
 By this approximation and direct integration, we obtain the approximate values of $\lambda_{ij}\;(i,j=1,2,3)$  as given by \eqref{lambdaforspheroid} in the main text.

\quad\\
\quad\\
\quad\\
\subsection*{Acknowlegement}
The authors thank Prof. Jian  Song for insightful discussions.

\bibliography{POFpaper}

\end{document}